%% file: quintessence.tex
\documentclass[letterpaper,12pt]{article}
\usepackage[DIV12]{typearea}
\usepackage{longtable}
\usepackage{booktabs,colortbl}
\usepackage{multirow}
\usepackage{amsmath,amsbsy,amstext,amsthm}
\usepackage{amssymb}
\usepackage{bbm}
\usepackage[utf8]{inputenc}
\usepackage{pdflscape}
\usepackage{xspace}
\usepackage[caption=false]{subfig}
\usepackage{nicefrac}
\usepackage{graphicx}
\usepackage{cite}  
\usepackage[small]{caption2}
\usepackage{mathtools}
\usepackage{nccfoots}
\usepackage{tikz}
\usepackage{enumerate}
\usepackage{dsfont}
\usepackage{transparent}
\usepackage{overpic}
\usepackage{import}
\usepackage[nolist]{acronym}

\usepackage[english]{babel}
\usepackage[bookmarks]{hyperref}
\hypersetup{
    urlcolor=NavyBlue,
    citecolor=NavyBlue,
    linkcolor=NavyBlue,
}%
\usepackage{array}
\usepackage{url} 
\usepackage{slashed}
\usepackage{multicol}
\usepackage{blkarray}
\usepackage{amsfonts}
\usepackage{amssymb}
  \usepackage{empheq}
  \usepackage{tcolorbox}
\usepackage{colortbl}
 \usepackage{float}
 \usepackage[normalem]{ulem}
\usepackage[utf8]{inputenc}
\RequirePackage{color}

\usetikzlibrary{calc,positioning}
\usepackage[textsize=tiny,backgroundcolor=yellow]{todonotes}

\usepackage[all,cmtip]{xy}
\newlength{\xywd}
\newcommand{\xyrightarrow}[2][]{%
  \sbox{0}{$\scriptstyle#1$}%
  \xywd=\wd0
  \sbox{0}{$\scriptstyle#2$}%
  \ifdim\wd0>\xywd \xywd=\wd0 \fi
  \xymatrix@C\dimexpr\xywd+1em\relax{{}\ar[r]^{#2}_{#1}&{}}%
}

\renewcommand{\thefootnote}{\alph{footnote}}

\DeclareMathOperator{\re}{Re}
\DeclareMathOperator{\im}{Im}

\newcommand{\rep}[1]{\ensuremath\boldsymbol{#1}}
\newcommand{\crep}[1]{\ensuremath\overline{\boldsymbol{#1}}}
\newcommand{\Z}[1]{\ensuremath{\mathds{Z}_{#1}}} % z_N ->\Z{N}
\newcommand{\SO}[1]{\ensuremath{\mathrm{SO}(#1)}}
\newcommand{\SU}[1]{\ensuremath{\mathrm{SU}(#1)}}
\newcommand{\SL}[1]{\ensuremath{\mathrm{SL}(#1)}}

\newcommand{\U}[1]{\ensuremath{\mathrm{U}(#1)}}
\newcommand{\E}[1]{\ensuremath{\mathrm{E}_{#1}}}
\newcommand{\e}{\mathrm{e}}
\newcommand{\I}{\mathrm{i}}
\newcommand{\Id}{\mathbbm{1}}

\newcommand{\x}{\ensuremath{\times}}
\newcommand{\vev}[1]{\ensuremath{\langle{#1}\rangle}}

\usepackage{xcolor}
\definecolor{darkgreen}{HTML}{109930}
\definecolor{pink}{rgb}{0.858, 0.188, 0.478}
\definecolor{redM}{RGB}{206, 0, 0}

\advance \headheight by 3.0truept       % for 12pt mandatory...
\RequirePackage{color}
\definecolor{auburn}{rgb}{0.43, 0.21, 0.1}
\definecolor{green}{rgb}{0.0, 0.5, 0.0}
\definecolor{darkviolet}{rgb}{0.58, 0.0, 0.83}
\definecolor{mediumviolet-red}{rgb}{0.78, 0.08, 0.52}
\def\be{\begin{equation}}
\def\ee{\end{equation}}
\def\bea{\begin{eqnarray}}
\def\eea{\end{eqnarray}}

\hypersetup{
   pdftitle = {Rolling with modular symmetry: quintessence and de Sitter in heterotic orbifolds},
   pdfauthor = {Gordillo-Ruiz, Hernandez-Segura, Portillo-Castillo, Ramos-Sanchez, Zavala}
}

\usepackage[toc,page]{appendix}

\newcommand{\diag}{\ensuremath{\text{diag}}} 
\usepackage{cleveref}

\begin{document}

% %%%%%%%ACRONYMS%%%%%%%%%%%%%%%
\begin{acronym}
         \acro{BU}{bottom-up}
         \acro{FI}{Fayet--Iliopoulos}
         \acro{IR}{infrared}
         \acro{SM}{standard model}
         \acro{TD}{top-down}
         \acro{UV}{ultraviolet}
         \acro{VEV}{vacuum expectation value}
         \acro{VVMF}{vector-valued modular form}
         \acro{dS}{de Sitter}
         \acro{adS}{anti-de Sitter}
         \acro{DE}{dark energy}
         \acro{EFT}{effective field theory}
         \acro{4D}{4-dimensional}
         \acro{CPL}{Chevallier-Polarski-Linder}
         \acro{LEEFT}{low-energy effective field theorie}
         \acro{GAP}{Groups, Algorithms, and Programming}
         \acro{SM}{Standard Model of particle physics}
         \acro{CFT}{conformal field theory}
         \acro{FLRW}{Friedmann--Lema\^itre--Robertson--Walker}
         \acro{GS}{Green--Schwarz}
         \acro{DDE}{{\em dynamical dark energy}}
         \acro{BBN}{big bang nucleosynthesis}
\end{acronym}

\begin{titlepage}

\vspace*{1.0cm}

\begin{center}
{\Large\textbf{Rolling with Modular Symmetry: \\[3mm]
Quintessence and de Sitter in Heterotic Orbifolds}}

\vspace{1cm}
\textbf{Hansel Gordillo--Ruiz}$^1$, 
\textbf{Miguel Hern\'andez--Segura}$^1$, \\
\textbf{Ignacio Portillo--Castillo}$^{1,2}$,
\textbf{Sa\'ul Ramos--S\'anchez}$^1$
and
\textbf{Ivonne Zavala}$^3$
\Footnote{*}{%
\href{mailto:hanselgordillo@estudiantes.fisica.unam.mx ;hernandez@estudiantes.fisica.unam.mx;iportillo@uach.mx;ramos@fisica.unam.mx;e.i.zavalacarrasco@swansea.ac.uk}{\tt Electronic addresses} 
}

$^1$\textit{\small Instituto de F\'isica, Universidad Nacional Aut\'onoma de M\'exico, Cd.~de M\'exico\\ C.P.~04510, M\'exico}\\
$^2$\textit{\small Facultad de Ingenier\'ia, Universidad Aut\'onoma de Chihuahua, Chihuahua \\ C.P. 31125, M\'exico.}\\
$^3$\textit{\small Centre for Quantum Fields and Gravity, Physics Department, \\Swansea University SA2 8PP, UK}
\end{center}

\vspace{1.5cm}

\begin{abstract}
Modular invariance is a fundamental symmetry in string compactifications, constraining both the structure of the effective theory and the dynamics of moduli and matter fields. It has also gained renewed importance in the context of swampland conjectures and, independently, flavour physics.
We investigate a modular-invariant scalar potential arising from heterotic orbifolds, where the flavour structure and moduli dynamics are jointly shaped by the underlying geometry. Focusing on a string-inspired, two-moduli truncation, we uncover a rich vacuum structure featuring anti-de Sitter minima and unstable de Sitter saddle points. We identify  large regions in moduli space
supporting multifield hilltop quintessence consistent with observations. All solutions satisfy refined swampland de Sitter bounds. Our results illustrate how modular symmetry can guide the construction of controlled, string-motivated quintessence scenarios within consistent effective theories.
\end{abstract}

\end{titlepage}

\newpage
\setcounter{footnote}{0} 
\renewcommand{\thefootnote}{\arabic{footnote}}

 \tableofcontents

%%%%%%%%%%%%%%%%%%%%%%%%%%%%%%%%%%%%%%%%%%%%%%%%%%%%%%%%%%%%%%%%%%%%%%%%%%%%%%%%%%%%%%%%%%%%%%%%%%%%%%%%%%%%%%%%%%%%%%%%%%%%%%%%%%%%%%%%%

\section{Introduction}
Heterotic orbifolds offer a remarkably fertile framework for deriving four-dimensional effective theories from string theory, combining realistic gauge structures with a deep intertwining of matter and moduli sectors~\cite{Bailin:1999nk,Lebedev:2006kn,Nilles:2011aj,Olguin-Trejo:2018wpw,Baur:2019iai,Ramos-Sanchez:2024keh}. In these models, compactifying the \E8\x\E8 heterotic string on a toroidal orbifold $ \mathds{T}^6 / P$, where $P$ is a discrete point group, yields chiral spectra, gauge symmetry breaking, and localized matter at orbifold fixed points. Crucially, such constructions inherit modular symmetries, remnants of target-space duality, which act non-trivially on both moduli and matter fields.

Far from being a residual geometric artifact, modular symmetry plays an active role in shaping effective theories from string compactifications. It constrains the structure of Yukawa couplings, governs the form of non-perturbative superpotentials, and serves as a guiding principle for constructing and validating complete \acp{LEEFT}.\footnote{Modular invariance was explicitly used to construct the full non-linear supergravity \ac{LEEFT} in concrete heterotic orbifold models in~\cite{Parameswaran:2010ec}; a similar principle guided the derivation of anti-D3 uplifted flux compactifications in~\cite{GarciadelMoral:2017vnz}.}

More recently, modular symmetry has entered the stage of the swampland program (see~\cite{vanBeest:2021lhn,Grana:2021zvf} for recent reviews). In~\cite{Olguin-Trejo:2018zun}, the unstable \ac{dS} vacua found in explicit heterotic orbifolds~\cite{Parameswaran:2010ec} were shown to satisfy the refined de Sitter swampland conjecture~\cite{Garg:2018reu,Ooguri:2018wrx}, which imposes a universal bound on scalar potentials in consistent quantum gravities:
\begin{equation}\label{eq:dSC}
 \frac{\sqrt{\nabla^i V\nabla_i V}}{V}  \geq \frac{c}{M_{\rm Pl}}  \, \qquad \text{or}\qquad \frac{{\rm min}(\nabla^i\nabla_j V)}{V} \leq -\frac{c'}{M_{\rm Pl}^2}\,,
\end{equation}
where ``min()'' denotes the minimal eigenvalue and $c$ and $c'$ some  ${\cal O}(1)$  positive constants. 
Subsequent work~\cite{Gonzalo:2018guu} further tested these constraints using modular-invariant potentials, again finding \ac{dS} saddle points and maxima (but no \ac{dS} minima) in line with~\cite{Parameswaran:2010ec,Olguin-Trejo:2018zun}  and swampland expectations. Related developments have pointed to a modular-invariant formulation of the species scale~\cite{vandeHeisteeg:2023ubh,Castellano:2023aum,vandeHeisteeg:2023dlw} and its implications for entropy bounds and effective field theory cutoffs.

In parallel, modular symmetry has re-emerged in cosmology. In the early universe, modular-invariant inflation was proposed~\cite{Schimmrigk:2014ica, Schimmrigk:2016bde,Schimmrigk:2021tlv,Kobayashi:2016mzg,Abe:2023ylh,King:2024ssx,Ding:2024neh}, and recently revived in the context of \(\alpha\)-attractors and swampland constraints~\cite{Casas:2024jbw,Aoki:2024ixq,Kallosh:2024pat,Kallosh:2024ymt}. In the late universe, the latest results from the Dark Energy Spectroscopic Instrument (DESI) \cite{DESI:2024kob,DESI:2024mwx,DESI:2025zpo,DESI:2025zgx,Lodha:2025qbg} 
and the Dark Energy Survey (DES)~\cite{DES:2024jxu,DES:2025bxy} suggest that the current acceleration may be due to a \ac{DDE}. One popular possibility for such \ac{DDE} is quintessence~\cite{Ratra:1987rm,Peebles:1987ek,Caldwell:1997ii}, described by a slowly rolling scalar.

Despite its challenges, which include phantom crossings and the unresolved cosmological constant problem, \ac{DDE} may offer a more natural string-theoretic origin than a positive cosmological constant. String-based models of exponential or hilltop quintessence have begun to emerge~\cite{Andriot:2024jsh, Bhattacharya:2024hep,Bhattacharya:2024kxp}, and recent works~\cite{Bernardo:2022ztc,Anchordoqui:2025epz,Bedroya:2025fwh} have pushed this dialogue between cosmology and the swampland further.

At the same time, a different motivation for modular symmetry has taken root: flavour. The proposal of~\cite{Feruglio:2017spp} that finite modular groups might underlie lepton flavour structures has found a natural setting in heterotic orbifolds~\cite{Baur:2024qzo}. 
Recent studies have shown that modular transformations give rise to modular flavour symmetries~\cite{Baur:2019kwi,Baur:2019iai}, which combine naturally with other moduli-independent symmetries of heterotic orbifolds, leading to the emergence of so-called eclectic flavour symmetries~\cite{Nilles:2020nnc,Nilles:2020gvu,Baur:2024qzo}. These symmetries place strong constraints on the \ac{LEEFT} of the compactifications~\cite{Lauer:1989ax,Lauer:1990tm,Nilles:2020kgo}, enabling both the reproduction of experimental observations and the formulation of concrete predictions in flavour physics~\cite{Baur:2022hma} via straightforward breaking patterns~\cite{Baur:2021bly}.

Flavour physics and moduli dynamics thus meet at a natural intersection in heterotic orbifolds. And yet, the interplay between these sectors, especially in the context of cosmological applications, remains largely unexplored. Our aim in this work is to take a step toward bridging this gap.

\bigskip

In this paper, we present a string-motivated model in which modular symmetry governs both the scalar potential for dynamical moduli and the flavour structure of matter fields. Our setup is based on the $\mathds T^6/\Z{6}-\text{II}$ orbifold, whose geometry includes a $ \mathds{T}^2 /\Z3$ sector. This naturally gives rise to the finite modular group $\Gamma_3' \cong T'$, which acts as a non-Abelian flavour symmetry on twisted-sector fields localised at fixed points. 

Building on the explicit heterotic orbifold construction of~\cite{Parameswaran:2010ec}, we construct a modular-invariant supergravity theory incorporating both double gaugino condensation and modular Yukawa couplings\footnote{In the recent work \cite{Leedom:2022zdm}, heterotic toroidal orbifold vacua for the overall K\"ahler modulus and the dilaton were investigated. That study incorporated non-perturbative corrections to both the superpotential and the K\"ahler potential, while preserving modular invariance. However, Yukawa couplings were not included, and only a single gaugino condensate was considered.}. We focus on scenarios with two dynamical moduli, the dilaton and a K\"ahler modulus; other moduli and matter fields are considered to develop \acp{VEV} while preserving supersymmetry at the compactification scale. This leads to a modular-invariant scalar potential suitable for studying both vacuum structure and cosmological dynamics.

Our results reveal a landscape populated by \ac{adS} minima and \ac{dS} saddle points\footnote{Recent work on unstable \ac{dS} solutions in string theory has appeared in \cite{{Chen:2025rkb,Andriot:2021rdy,Andriot:2024cct,Parameswaran:2024mrc,ValeixoBento:2025yhz}}.}
consistent with swampland expectations as well as multifield rolling trajectories supporting hilltop quintessence\footnote{For recent work on multifield quintessence models see \cite{Cicoli:2020noz,Akrami:2020zfz,Anguelova:2021jxu, Anguelova:2023dui, Eskilt:2022zky, Seo:2024qzf,Licciardello:2025fhx}.}. In particular, we find large regions
in moduli space where rolling trajectories along a single tachyonic direction cluster. While its full physical origin remains to be understood, its emergence may signal hidden modular patterns in the potential. 

This work also realises the hilltop quintessence scenario of~\cite{Olguin-Trejo:2018zun} in a fully modular setting, unifying string-derived flavour structure with rolling scalar dynamics. In doing so, it offers a new template for embedding late-time cosmology in UV-complete frameworks.

\medskip

The paper is structured as follows. In \Cref{sec:2}, we review modular symmetries in heterotic orbifolds and their consequences for the effective theory. \Cref{sec:numresults} presents the numerical landscape of vacua and dynamical solutions. Our analysis of multifield quintessence and cosmological trajectories appears in \Cref{sec:quintessence}, followed by conclusions and future outlook in \Cref{sec:conclusions}. In  \Cref{App:concrete_model} we provide the details of a $\mathds T^6/\Z{6}-\text{II}$ orbifold  model used as a basis of our study. 
In the following, unless stated otherwise, we adopt the reduced Planck units with $M_\mathrm{Pl}=1$.

\bigskip

%%%%%%%%%%%%%%%%%%%%%%%%%%%%%%%%%%%%%%%%%%%%%%%%%%%%%%%%%%%%%%%%%%%%%%%%%%%%%%%%%%%%%%%%%%%%%%%%%%%%%%%%%%%%%%%%%%%%%%%%%%%%%%%%%%%%%%%%%%
\input{Section2}
\input{numericalsearch}
\input{results}   
%%%%%%%%%%%%%%%%%%%%%%%%%%%%%%%%%%%%%%%%%%%%%%%%%%%%%%%%%%%%%%%%%%%%%%%%%%%%%%%%%%%%%%%%%%%%%%%%%%%%%%%%%%%%%%%%%%%%%%%%%%%%%%%%%%%%%%%%%%
\section{Discussion and Outlook}
\label{sec:conclusions}

Heterotic orbifolds are among the simplest string constructions that can reproduce many features of particle physics, thanks to their rich symmetry structure. In particular, modular symmetries naturally appear in these models as inherited from the toroidal structure of the compact space. Such symmetries have recently attracted wide attention both in the context of flavour physics, where they serve as instrumental restrictions to reach predictions, and in the swampland program, where they provide a guiding principle for formulating quantum gravity constraints. Thus, heterotic orbifolds provide a natural arena for connecting particle phenomenology, swampland constraints, and cosmological dynamics. 

In this work, we have shown that heterotic orbifolds combine these ingredients into a coherent framework. 
As a working example, we studied a two-dimensional $\mathds T^2/\Z{3}$ orbifold sector of a heterotic orbifold compactification without Wilson lines. In this case, the \SL{2,\Z{}} modular symmetry governing the dynamics of the toroidal modulus $\tau$ is realised as a $\Gamma_3' \cong T'$ finite modular symmetry of the effective action. The coupling strengths among matter superfields are then described by \acp{VVMF}, which depend solely on $\tau$ and build $T'$ representations. Exotic matter associated with hidden non-Abelian gauge sectors is typically decoupled, but their gaugino condensates contribute modular-invariant terms to the effective action, coupling $\tau$ and the dilaton $S$. Other moduli can be stabilised supersymmetrically, as we assume to be the case. 
The resulting setup, based on the restrictions of the modular $T'$ and gaugino condensates, drives the dynamics of the lightest complex moduli $\tau$ and $S$, which may control aspects of cosmology.

We first performed in \Cref{sec:numresults} a systematic exploration of the critical points of the modular-invariant potential, by varying the integers $m_a$ and $n_a$ in the non-perturbative superpotential. 
Our extensive numerical search uncovered several interesting features for the structure of these critical points, which we classified according to our scheme outlined in \Cref{fig:Tree}:
\begin{itemize}
    \item No \ac{dS} (meta)-stable vacua were found, in line with the \ac{dS} swampland conjecture and previous work \cite{Parameswaran:2010ec}.
    \item The solutions fall into two categories: \emph{unstable \ac{dS} critical points} and \emph{stable and unstable \ac{adS} vacua}. 
    \item The unstable saddles satisfy the refined \ac{dS} conjecture and exhibit a non-trivial structure in their tachyonic directions, including an interesting ``penacho'' distribution along axionic directions shown in \Cref{fig:Tau} and distinctive patterns in the dilaton plane as shown in \Cref{fig:S}. These may simply be coincidences, but determining their relevance requires further study.
\end{itemize}

We then studied the cosmological implications of this setup in \Cref{sec:quintessence}. These are twofold. First, unstable \ac{dS} saddles provide natural realisations of multifield hilltop quintessence models, which satisfy swampland constraints. In particular, we showed that axionic combinations of the moduli can drive slow-roll dynamics compatible with the recent observations hinting at a dynamical form of dark energy. Our explicit example yields an equation-of-state evolution consistent with DESI within $3\sigma$. This suggests heterotic orbifolds as concrete string-based realisations of quintessence, while satisfying swampland constraints. Of course, a full cosmological analysis, including perturbations and couplings to matter, remains to be performed. We stress that this framework does not eliminate the well-known challenges of quintessence, such as fine-tuning, fifth-force bounds, variations of fundamental constants, and the cosmological constant problem itself. For example, as shown in \Cref{table:parametersAdS}, the saxion turns out to be very light, which could potentially lead to conflicts with fifth-force constraints, a question that merits further investigation. However, our model provides a natural and \ac{UV}-complete setting in which these challenges can be systematically addressed. 

Second, the stable \ac{adS} vacua we identified offer further insight into the structure of the landscape. In these solutions, the Kaluza-Klein length scale is generically negligible compared with the \ac{adS} curvature scale, thereby satisfying the conjecture of scale separation in quantum gravity. 

Several avenues remain to be explored and we plan to come back to this in future work. Firstly, a systematic investigation of the ``penacho'' structure and other statistical patterns in the distribution of tachyonic modes will allow us to determine whether they have a modular or geometric origin. Our setup provides a robust proof of principle, but extending the analysis to include more realistic compactifications with additional moduli and matter fields remains an important next step. Furthermore, exploiting the modular symmetries inherent in heterotic orbifolds to probe quantum gravity conjectures more sharply seems a natural step forward as well. 
On the cosmological side, it will be interesting to develop a multifield generalisation of the DSCh hilltop parametrisation of the equation of state, enabling a broader comparison between string-based quintessence and data. 
Furthermore, although we have studied here a model displaying thawing-like behaviour, where the equation of state $w_{\rm DE}$ grows from $-1$ towards larger values, current analyses also hint at a possible crossing of the phantom divide~\cite{DESI:2025zgx,DESI:2025fii}, which motives further research to realise this non-standard quintessence scenario in string constructions. Also, exploring couplings between dark energy and dark matter within this framework, where an effective phantom-like behaviour may arise \cite{Das:2005yj} is left for future work.  

Heterotic orbifolds offer a unifying framework in which modular symmetries, swampland conjectures, particle phenomenology, and cosmology intersect. Our results demonstrate that the landscape of extrema is both structured and phenomenologically rich. Rather than being pathologies to avoid, unstable saddles emerge as natural candidates for quintessence within a \ac{UV}-complete setting, which deserve further study.

\vspace{-4mm}
\subsection*{Acknowledgments}
It is a pleasure to thank  Bruno V.~Bento and Timm Wrase for useful discussions and Susha Parameswaran for comments on an earlier version of the manuscript. 
This work was partially supported by UNAM-PAPIIT grants IN113223 and IN117226, and the Marcos Moshinsky Foundation. IZ is partially funded by the STFC grants
ST/T000813/1 and ST/X000648/1. For the purpose of open access, the authors have applied a Creative Commons Attribution license to any Author Accepted Manuscript version arising. Research Data Access Statement: No new data were generated for this manuscript.

%\newpage
%%%%%%%%%%%%%%%%%%%%%%%%%%%%%%%%%%%%%%%%%%%%%%%%%%%%%%%%%%%%%%%%%%%%%%%%%%%%%%%%%%%%%%%%%%%%%%%%%%%%%%%%%%%%%%
\appendix

\input{appendixA}

%%%%%%%%%%%%%%%%%%%%%%%%%%%%%%%%%%%%%%%%%%%%%%%%%%%%%%%%%%%%%%%%%%%%%%%%%%%%%%%%%%%%%%%%%%%%%%%%%%%%%%%%%%%%%%

\newpage

{\small
\addcontentsline{toc}{section}{References}
%\bibliographystyle{utphys}
%\bibliography{Orbifold}

\providecommand{\href}[2]{#2}\begingroup\raggedright\endgroup

}
\end{document}

%% file: Section2.tex
\section{Heterotic orbifold compactifications}
\label{sec:2}

Toroidal heterotic orbifolds are a class of compactifications in which the extra six dimensions of a heterotic string are taken to be a quotient of a six-torus $\mathds{T}^6$ by a discrete symmetry group. Explicitly, the compact space is defined as
\begin{equation}
\mathcal{O}_6 ~=~ \mathds{T}^6 / P\,,
\end{equation}
where $P$ is a finite group of discrete isometries of the torus, known as \emph{point group}. To be compatible with the toroidal structure, $P$ must act crystallographically on the underlying lattice $\Lambda$ defining $\mathds{T}^6$, which we consider factorisable: $\mathds{T}^6=\mathds{T}^2\times \mathds{T}^2\times \mathds{T}^2$. The point group acting on $\mathds{T}^6$ is generated by \SO6 rotations called twists, which in complex coordinates can be written as $\diag(\e^{2\pi\I v_1},\e^{2\pi\I v_2},\e^{2\pi\I v_3})$, where $v=(v_1,v_2,v_3)^\mathrm{T}$ is called twist vector.

It is often useful to instead define the orbifold in terms of its \emph{space group} $\mathcal{S}$, which, in the absence of roto-translations, is given by the semidirect product
\begin{equation}
\mathcal{S} ~=~ P \ltimes \Lambda\,.
\label{eq:spacegroup}
\end{equation}
In this formulation, the orbifold is equivalently defined as a quotient of flat space by the space group,
\begin{equation}
\mathcal{O}_6 ~=~ \mathds{R}^6 / \mathcal{S}\,.
\label{eq:defOrbifoldinS}
\end{equation}
An important quality of an orbifold is that it is flat everywhere, but at a finite number of curvature singularities, which are those geometric points in $\mathds R^6$ that are left fixed under the action of particular elements of $\mathcal S$.

The consistency of the heterotic strings demands the embedding of the orbifold $\mathcal{O}_6$ into the 16 gauge degrees of freedom. This can be done via a 16-dimensional shift vector $V$ acting on the left-moving momenta. Focusing on the \E8\x\E8 heterotic string, $V$ shifts the momenta in the gauge lattice $\Lambda_{\E8\x\E8}$ and thereby breaks down the gauge group \E8\x\E8 to a 4-dimensional subgroup, which includes in many cases the gauge group of the \ac{SM}~\cite{Buchmuller:2006ik,Lebedev:2006kn,Nilles:2014owa,Olguin-Trejo:2018wpw,Parr:2019bta,Baur:2022hma}. Twist and shift vectors must satisfy {\em modular invariance} constraints on the string worldsheet (see e.g.~\cite{Bailin:1999nk,Ploger:2007iq,Ramos-Sanchez:2008nwx,Vaudrevange:2008sm,Choi:2020dws} for more details).

The massless matter spectrum of a heterotic orbifold comprises two kinds of string states: {\it untwisted} and {\it twisted fields}. Untwisted matter fields are associated with those closed strings of the original string theory  that are left invariant by the orbifold action. They are free to move in the bulk, the whole compact space. Twisted matter fields are related to strings that close only because of the (twist) action of the orbifold and are attached to the geometric fixed points of the orbifold. They all build various gauge representations, which, in realistic cases, include those of the \ac{SM} quarks and leptons. Since the twisted states are linked to orbifold singularities, the replication of families may find an explanation in the number of fixed points of the orbifold.

%%%%%%%%%%%%%%%%%%%%%%%%%%%%%%%%%%%%%%%%%%%%%%%%%%%%%%%%%%%%%%%%%%%%%
\subsection{Target space modular symmetry}
\label{sec:Modular}

The spectrum of states in an orbifold string compactification is invariant under 
 target-space modular symmetry arising from T-duality \cite{Ferrara:1989bc,Font:1990nt,Cvetic:1991qm}, acting as discrete transformations on the moduli. Focusing on factorisable six-tori, each $\mathds T^2$ includes two moduli: a complex structure $U$ and a K\"ahler modulus $\tau$. Under the orbifold action, the complex structure is geometrically fixed in all cases but $\mathds T^2/\Z2$~\cite{Nilles:2020gvu}. In contrast, the K\"ahler modulus $\tau$ is free and subject to a $\SL{2, \Z{}}$ target-space modular symmetry, which  acts as\footnote{In the $\mathds T^2/\Z2$ case, the target-space duality group is $\SL{2, \Z{}}_\tau \x  \SL{2, \Z{}}_U$ for null Wilson lines. In this work, we focus only on K\"ahler moduli as the complex structure is geometrically fixed in the case we shall consider (see~\cite{Paradisi:2008qh} where the complex structure was also kept free).} 
\begin{equation}
\label{eq:modt}
    \tau ~\xmapsto{~\gamma~}~ \gamma\tau ~:=~
    \frac{a \tau + b}{c\tau + d}\, , \qquad \gamma \in \Gamma\, ,
\end{equation}
where 
\begin{equation}
   \Gamma := \SL{2,\Z{}} ~=~ \left\{
    \gamma:=\begin{pmatrix}
        a & b \\
        c & d 
    \end{pmatrix} ~\vline~
    \; a, b, c, d \in \Z{},\quad ad-bc = 1
    \right\}\,. 
\end{equation}

\begin{figure}[t]
    \centering
    \includegraphics[width=0.6\linewidth]{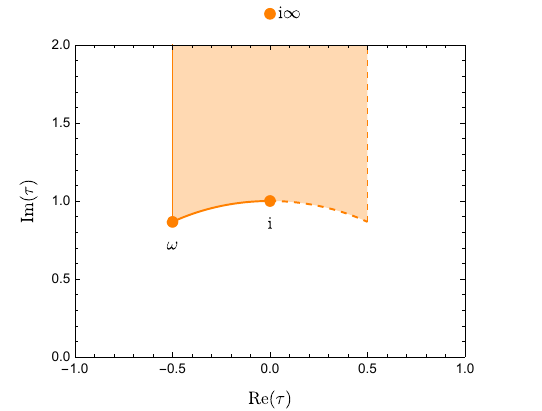}
    \caption{\label{fig:dominiofundamental} 
    Fundamental domain of  $\SL{2,\Z{}}$ for the modulus $\tau$. }
\end{figure}

\noindent The generators of $\Gamma$ can be chosen to satisfy the presentation 
\begin{equation}
    \left\langle~ \mathrm S\,, \mathrm T ~\vline~ (\mathrm{ST})^3 = \mathrm{S}^4 = \Id\,,\mathrm{T\,S}^2 = \mathrm{S}^2\mathrm{T} ~\right\rangle
\end{equation}
and are frequently represented by
\begin{equation}
    \mathrm{S} ~=~ \begin{pmatrix}
        0 & 1 \\
        -1 & 0
    \end{pmatrix} \qquad \text{and}\qquad
    \mathrm{T} ~=~ \begin{pmatrix}
        1 & 1 \\
        0 & 1
    \end{pmatrix}\, .
\end{equation}
We see that $\tau$ transforms identically under both
$\gamma$ and $-\gamma$, which implies
that $\tau$ only transforms non-trivially under $\mathrm{PSL}(2, \Z{})=\Gamma/\Z2$.
Hence, the modulus $\tau$ takes values only in the upper half-plane
\begin{equation}
    \mathcal{H} ~=~ \left\{\tau\in \mathds{C} \;\;\vline \;\;\mathrm{Im}(\tau)>0\right\}\, .
\end{equation}
Further, the modular symmetry restricts the values of the modulus to the fundamental domain of $\Gamma$, which corresponds to the region illustrated in \Cref{fig:dominiofundamental}. The modular transformations $\Gamma$ leave three
fixed points in the fundamental domain:
\begin{itemize}
    \item $\tau = \mathrm{i}$, fixed by $\mathrm{S}$;
    \item $\tau = \omega := \e^{2\pi\mathrm{i}/3}$,
     fixed by $\mathrm{ST}$; and
    \item $\tau= \mathrm{i}\infty$, fixed by $\mathrm{T}$.
\end{itemize}

There are interesting modular groups that can be obtained from the quotients of $\Gamma=\SL{2,\Z{}}$ and its principal congruence subgroups $\Gamma(N)$, which are defined as 
\begin{equation}
\Gamma(N)~:=~\left\{ \gamma \in \Gamma ~|~ \gamma = \Id\, \bmod N \right\}. 
\end{equation} 
The quotient $\Gamma/\Gamma(N) =:\Gamma'_{N}$ is a so-called finite modular group of level $N$.

In general, associated with any congruence modular group, there are holomorphic functions of $\tau\in\mathcal{H}$ of level $N$, known as modular forms, which change under modular transformations according to
\begin{equation}
\label{eq:modtrafof}
    f_i(\tau) ~\xmapsto{~\gamma~}~ f_i(\gamma\,\tau) ~=~ (c\,\tau+d)^{n_{f_i}}f_i(\tau)\,, \qquad \gamma =\begin{pmatrix}
    a & b \\
    c & d
\end{pmatrix}\in\Gamma(N),
\end{equation}
where $n_{f_i}\in\mathds{N}$ denotes the so-called modular weight of $f_i$ and $(c\tau+d)^{n_{f_i}}$ is known as automorphy factor. Note that $\gamma$ in \Cref{eq:modtrafof} is an element of $\Gamma(N)$ and not of $\Gamma$. Indeed, under $\gamma\in\Gamma$ modular forms of level $N$ and identical weight $n_F$ transform non-trivially within finite-dimensional vector subspaces of modular forms. Considering one of these subspaces of dimension $s$, the vector $\hat F^{(n_F)}_{\rep s}(\tau)=(f_{1}(\tau),\ldots, f_{s}(\tau))^{\mathrm T}$ is a so-called \ac{VVMF}~\cite{Liu:2021gwa} and transforms according to
\begin{equation}
\label{eq:VVMF}
    \hat F^{(n_F)}_{\rep s}(\gamma\,\tau) ~\xmapsto{~\gamma~}~ 
    \hat F^{(n_F)}_{\rep s}(\gamma\,\tau)~=~(c\,\tau+d)^{n_{F}}\,\rho_{\rep s}(\gamma)\,\hat F^{(n_F)}_{\rep s}(\tau),\qquad \gamma\in\Gamma,
\end{equation}
with $\rho_{\rep s}(\gamma)$ a unitary irreducible $s$-dimensional representation of $\gamma$ in the finite group $\Gamma'_{N}$. Similar properties hold for other finite modular groups resulting from the quotients of $\Gamma$ and any of its normal subgroups (see e.g.~\cite{Ding:2023ydy,Arriaga-Osante:2023wnu,Arriaga-Osante:2025ppz} for examples).

On the other hand, since heterotic orbifolds are based on toroidal compactifications, it is natural to expect that some of these transformations arise naturally as target-space modular symmetries. In fact, it turns out that heterotic orbifolds exhibit finite modular groups as flavor symmetries of the \ac{LEEFT}~\cite{Baur:2024qzo}. To uncover the modular symmetries that arise in heterotic orbifolds, 
it is useful to switch to the Narain description of the compactification of the heterotic string over an orbifold~\cite{Narain:1985jj}.  
In this formalism, the outer automorphisms of the Narain space group\footnote{Not to be confused with the space group $\mathcal{S}$ in~\Cref{eq:spacegroup}. The Narain space group codifies both the geometric and the gauge degrees of freedom, providing a richer description of the compactification, see e.g.~\cite{GrootNibbelink:2017usl}.} 
encode the flavour symmetries of the \ac{LEEFT}~\cite{Baur:2019kwi,Baur:2019iai,Nilles:2020gvu}. These automorphisms can be classified either as purely translational or rotational. The rotational 
outer automorphisms of the Narain space group precisely build the target-space modular symmetries of the theory. One can further show that these symmetries are realised both by matter fields and their coupling strengths as finite modular groups. For instance, it is known that $\mathds{T}^2/\mathds{Z}_K$ orbifolds, $K=2,3,4,6$, lead respectively to the finite modular groups\footnote{We provide in brackets the GAP Id, given by the program  Groups, Algorithms, and Programming (GAP)~\cite{GAP4}. The first number is the order of the finite group, the second is a counter.} $(S_3\x S_3)\rtimes\Z4\cong[144,115]$, $\Gamma_3'\cong T'\cong[24,3]$, $2D_3\cong S_4'/(\Z2\x\Z2)\cong[12,1]$, and $\Gamma_6'\cong S_3\x T' \cong[144,128]$~\cite{Nilles:2020kgo,Baur:2024qzo}. Invariance under these target-space modular symmetries 
imposes non-trivial constraints on the form of the allowed background fields, the symmetry properties of the matter fields, their couplings, and the consistency of the effective theory.

Under the action of the modular group, untwisted and twisted matter fields display similar transformations to those of \acp{VVMF}. Explicitly,
\begin{equation}\label{eq:matterp}
    \Phi^{(n)} ~\xmapsto{~\gamma~}~ 
    \Phi^{(n)'} ~:=~ (c\tau + d)^n \, \rho_{\rep{r}}(\gamma) \, \Phi^{(n)}\,, \qquad \gamma\in\Gamma\,,
\end{equation}
where $(c\tau + d)^n$ is again an automorphy factor and $\rho_{\rep{r}}(\gamma)$ is an $r$-dimensional matrix representation of $\gamma$ in the finite modular group. 
Untwisted states transform as singlets of the finite modular group while twisted matter fields localised at the various orbifold singularities build $r$-dimensional field multiplets. Their modular weights $n$ are determined by the geometric properties of the associated strings~\cite{Ibanez:1992hc,Olguin-Trejo:2017zav}. While for untwisted fields they are either $-1$ or $0$, for twisted fields they possess various fractional values, which can be both positive and negative.

\subsubsection[\texorpdfstring{The $\mathds{T}^2/\Z3$ orbifold}{The T2/Z3 orbifold}]{\boldmath The $\mathds{T}^2/\Z3$ orbifold \unboldmath}
\label{sec:Z3orbifold}

In this work, we focus on the well-known example of the $\mathds{T}^2/\mathds{Z}_3$ orbifold, where the finite modular symmetry is $\Gamma_3'\cong T'$, and modular properties of the \ac{LEEFT} are well understood.

\begin{figure}[t]
  \centering
  \includegraphics[width=0.5\textwidth]{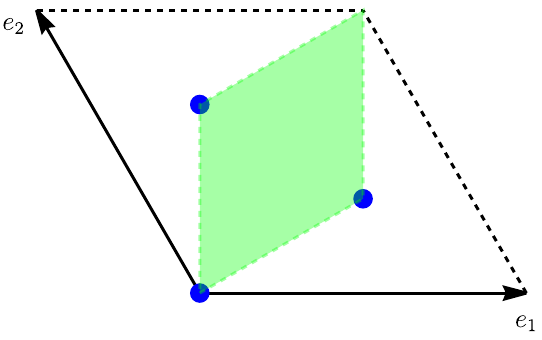}
  \caption{The $\mathds{T}^2/\Z3$ orbifold. The three inequivalent fixed points under the $\Z3$ action are shown as blue dots. 
  The green shaded region denotes the fundamental domain of the orbifold.}
  \label{fig:Z3orbifold}
\end{figure}

The $\mathds T^2/\Z3$ orbifold is produced by modding out a two-torus by a \Z3 isometry generated by the twist $\theta = \exp(2\pi \I/3)$. The resulting geometry is shown in \Cref{fig:Z3orbifold}. Points that are inequivalent on the torus are identified under the action of $\theta$, which leads to three orbifold fixed points. Further, the fundamental domain of the orbifold is only one third of the fundamental region of the torus.

Untwisted or bulk matter fields in four dimensions arise from the decomposition of the ten-dimensional gauge bosons $A^M$ of \E8\x\E8, with $M = 0, \dots, 9$. This decomposition depends on the directions wrapped by the orbifold:
\begin{itemize}
    \item Considering that $M=6,7$ correspond to the $\mathds{T}^2/\Z3$ orbifold, the associated compact dimensions give rise to untwisted scalars with modular weight $-1$, denoted by $\Phi^{(-1)}$.
    \item The remaining compact directions $M = 4,5,8,9$ yield the untwisted fields $\Phi^{(0)}$ with modular weight $0$.
\end{itemize}
These untwisted matter fields are trivial singlets under the finite modular group $T'$.

The three fixed points of $\mathds{T}^2/\Z3$ host twisted strings, which are localised at those singularities. These states are associated with either the first or second twisted sectors, depending on whether the corresponding strings close under $\theta$ or $\theta^2$. Further, the states may include additional oscillator excitations, which modify their specific modular weights. Hence, the available twisted matter fields are
\begin{itemize}
    \item $\Phi^{(-\nicefrac23)}$ and $\Phi^{(-\nicefrac53)}$ from the first twisted sector, and
    \item $\Phi^{(-\nicefrac13)}$,  $\Phi^{(\nicefrac23)}$,  $\Phi^{(-\nicefrac43)}$ and $\Phi^{(\nicefrac53)}$ from the second twisted sector.
\end{itemize}
The twisted matter states $\Phi^{(n)}$ include three massless fields each, which exhibit identical gauge quantum numbers.\footnote{Note that this multiplicity offers an opportunity to explain the observed family repetition in the \ac{SM}.} They can thus be considered flavour multiplets, with non-trivial transformations under the modular flavour symmetry $T'$. In the following, we discuss the details of these transformations and the properties of their couplings.

%%%%%%%%%%%%%%%%%%%%%%%%%%%%%%%%%%%%%%%%%%%%%%%%%%%%%%%%%%%%%%%%%%%%%
\subsubsection[\texorpdfstring{$T'$ transformations of couplings and twisted matter}{T' transformations of couplings and twisted matter}]{\boldmath $T'$ transformations of couplings and twisted matter\unboldmath}

$\Gamma_3'\cong T'\cong[24,3]$ is generated by $\mathrm S$ and $\mathrm T$, subject to the conditions 
\begin{equation}
    \mathrm S^4~=~\Id\,,\qquad 
    (\mathrm{ST})^{3}~=~\Id\,,\qquad 
    \mathrm S^2 \mathrm T=\mathrm{T S}^2\,,\qquad 
    \mathrm T^{3} ~=~ \Id\,,
\end{equation}
where the last relation renders the group finite.
In this case, there are two modular forms, $\hat{Y}_1(\tau)$ and $\hat{Y}_2(\tau)$, with lowest modular weight $n_Y=1$. They build a two-dimensional \ac{VVMF}, transforming as a $\rep2''$ of $T'$~\cite{Lu:2019vgm}, such that

\begin{equation}
\label{eq:modularforms}
    \hat{Y}^{(1)}_{\rep{2}''}(\tau) ~=~ 
    \begin{pmatrix}
        \hat{Y}_1(\tau)\\
        \hat{Y}_2(\tau)
    \end{pmatrix} ~:=~ 
    \begin{pmatrix}
        -3\sqrt{2} \frac{\eta^3(3\tau)}{\eta (\tau)}\\
        3\frac{\eta^3(3\tau)}{\eta (\tau)}+\frac{\eta^3(\tau/3)}{\eta (\tau)} \,  
    \end{pmatrix},
\end{equation}
with $\eta(\tau)$ the well-known Dedekind $\eta$ function, defined as
\begin{equation}\label{eq:etaq}
    \eta (\tau) ~=~ q^{\frac{1}{24}}\prod_{n=1}^\infty (1-q)^n ~=~ q^{\frac{1}{24}} \sum_{n=-\infty}^{\infty} (-1)^n q^{\frac12n(3n - 1)} \, , 
\end{equation}
where $q := \exp (2\pi \I \tau)$. From \Cref{eq:VVMF} we observe that $\hat{Y}^{(1)}_{\rep{2}''}$ transforms under $\gamma\in\SL{2,\Z{}}$ as
\begin{equation}
\label{eq:Ygeneraltransformation}
   \hat{Y}^{(1)}_{\rep{2}''}(\tau)~\xmapsto{~\gamma~}~ 
   \hat{Y}^{(1)}_{\rep{2}''}\left(\gamma\tau \right) ~=~ 
   \left(c\tau +d \right)^{1}
   \rho_{\rep{2}''}(\gamma) \hat{Y}^{(1)}_{\rep{2}''}(\tau)
\end{equation}
with $\rho_{\rep{2}''}(\mathrm{S})$ and $\rho_{\rep{2}''}(\mathrm{T})$ given by~\cite{Nilles:2020kgo}
\begin{equation}
\label{eq:2pprime}
    \rho_{\rep{2}''}(\mathrm{S}) ~:=~ -\frac{\I}{\sqrt{3}}\begin{pmatrix}
        1 & \sqrt{2}\\
        \sqrt{2} & -1
    \end{pmatrix}
    \qquad \text{and} \qquad \rho_{\rep{2}''}(\mathrm{T}) ~:=~\begin{pmatrix}
        \omega & 0\\
       0 & 1
    \end{pmatrix} \, . 
\end{equation}
\acp{VVMF} $\hat Y^{(n_Y)}_{\rep s}(\tau)$ with higher weights, $n_Y>1$, can be straightforwardly constructed from tensor products of $\hat{Y}^{(1)}_{\rep{2}''}$; e.g.\ $\hat{Y}^{(2)}_{\rep{3}}=\hat{Y}^{(1)}_{\rep{2}''}\otimes \hat{Y}^{(1)}_{\rep{2}''}$, where the singlet disappears because it is antisymmetric. A crucial observation in heterotic orbifolds (and other string constructions) is that the couplings among string fields are, in fact, modular forms~\cite{Erler:1992gt,Stieberger:1992vb}. Hence, since string couplings build non-trivial $T'$ representations, so should string matter multiplets $\Phi^{(n)}$ in order to arrive at a modular-invariant \ac{LEEFT}.

By using \ac{CFT} techniques, it is possible to uniquely determine the representations of twisted matter fields~\cite{Baur:2019kwi,Baur:2019iai}. In particular, as summarised in \Cref{tab:modularflavourcharges}, one finds that each multiplet of matter fields from the first twisted sector transforms as a $\rep2'\oplus\rep1$ of $T'$, while field multiplets from the second twisted sector transform as $\rep2''\oplus\rep1$~\cite{Baur:2024qzo}. The doublet representations of the $T'$ generators acting on these fields are given by \Cref{eq:2pprime} and
\begin{equation}\label{eq:2prime}
    \rho_{\rep{2}'}(\mathrm{S}) ~:=~ -\frac{\I}{\sqrt{3}}\begin{pmatrix}
        1 & \sqrt{2}\\
        \sqrt{2} & -1
    \end{pmatrix}
    \qquad \text{and} \qquad \rho_{\rep{2}'}(\mathrm{T}) ~:=~\begin{pmatrix}
        1 & 0\\
       0 & \omega^{2}
    \end{pmatrix} \, . 
\end{equation}

\begin{table}[]
    \centering
    \begin{tabular}{|c|c|c|c|c|c|c|c|c|}
    \hline
     \rowcolor{gray!30}
        \multicolumn{1}{|c}{sector:} & \multicolumn{2}{c|}{untwisted} & \multicolumn{2}{c|}{first twisted} & \multicolumn{4}{c|}{second twisted} \\
        \hline 
        string state  & $~\Phi^{(0)}~$ & $~\Phi^{(-1)}~$  & $\Phi^{(-\nicefrac23)}$ & $\Phi^{(-\nicefrac53)}$ & $\Phi^{(-\nicefrac13)}$ & $\Phi^{(\nicefrac23)}$ & $\Phi^{(-\nicefrac43)}$ & $\Phi^{(\nicefrac53)}$ \\
        \hline 
        $T'$ irrep    & \multicolumn{2}{c|}{$\rep1$}  & \multicolumn{2}{c|}{$\rep{2}'\oplus\rep{1}$} & 
        \multicolumn{4}{c|}{$\rep{2}''\oplus\rep{1}$} \\
       \hline
    \end{tabular}
    \caption{\label{tab:modularflavourcharges} Modular properties of matter multiplets of the $\mathds T^2/\Z{3}$ orbifold by sector~\cite{Baur:2024qzo}.}
\end{table}

%%%%%%%%%%%%%%%%%%%%%%%%%%%%%%%%%%%%%%%%%%%%%%%%%%%%%%%%%%%%%%%%%%%%%
\subsubsection[\texorpdfstring{Modular $T'$ from the $\Z6-\text{II}$ orbifold}{Modular T' from the Z6-II orbifold}]{\boldmath Modular $T'$ from the $\Z{6}-\text{II}$ orbifold \unboldmath}

There are six different \Z6 supersymmetric orbifolds in six dimensions~\cite{Fischer:2012qj}. Among them, the so-called $\Z{6}-\text{II}\, (1,1)$ orbifold features a factorised structure as $\mathds T^2/\Z6\otimes\mathds T^2/\Z3\otimes\mathds T^2/\Z2$, which has been shown to be phenomenologically fruitful (see e.g.~\cite{Kobayashi:2004ud,Kobayashi:2004ya,Buchmuller:2006ik,Parameswaran:2010ec}). This structure is particularly relevant for models of physics based on modular symmetries as the second torus offers a finite modular symmetry $\Gamma_3'\cong T'$, which together with a traditional flavour symmetry renders promising predictions for quarks and leptons~\cite{Baur:2022hma}. 

By e.g.\ using the \texttt{orbifolder}~\cite{Nilles:2011aj}, one can show that the matter spectrum of $\Z6-\text{II}\, (1,1)$ orbifolds offers both \ac{SM} matter and gauge singlets transforming as described in \Cref{tab:modularflavourcharges}. Most of these states are $\Phi^{(-\nicefrac23)}$ fields, whose couplings have been explicitly studied before~\cite{Nilles:2020kgo}. In particular, as we shall describe shortly, it is known that the modular forms $\hat Y_i(\tau)$ given in \Cref{eq:modularforms} appear in the trilinear couplings $\Phi^{(-\nicefrac23)}_\alpha\otimes\Phi^{(-\nicefrac23)}_\beta\otimes\Phi^{(-\nicefrac23)}_\gamma$ in the \ac{LEEFT} of these orbifolds, rendering the action modular invariant and thus providing a source for the (stabilising) potential of the modulus $\tau$. We shall base our study on these properties.

In accordance with the general framework of $\mathds{T}^2/\Z{K}$ orbifolds~\cite{Baur:2024qzo}, one might expect the associated \ac{LEEFT} of $\Z6-\text{II}\, (1,1)$ orbifolds to exhibit additional non-Abelian modular symmetries beyond $T'$. Interestingly, all semi-realistic models based on the $\Z6-\text{II}\,(1,1)$ orbifold incorporate Wilson lines~\cite{Lebedev:2008un,Olguin-Trejo:2018wpw}, which in many cases are associated with the $\mathds{T}^2/\Z2$ orbifold sector. Since these background fields explicitly break the modular symmetries of the orbifold sectors to which they are coupled~\cite{Bailin:1993ri,Love:1996sk}, no non-Abelian modular symmetry can arise from the third torus.
However, the $\mathds{T}^2/\Z6$ orbifold sector yields a $\Gamma_6'\cong S_3\x T'$ finite modular symmetry, under which (only) a few matter fields transform as doublets. For simplicity, since $T'$ appears as part of the modular symmetry of the K\"ahler modulus $\tau_1$ of $\mathds{T}^2/\Z6$ too, we shall assume that it can be stabilised by using the same mechanism we describe here; hence, we shall not discuss it explicitly in the following.

%%%%%%%%%%%%%%%%%%%%%%%%%%%%%%%%%%%%%%%%%%%%%%%%%%%%%%%%%%%%%%%%%%%%%
\subsubsection{Supersymmetric stabilisation of matter fields}
\label{sec:SUSYmodStab}

Prior to describing the specifics of the \ac{LEEFT} that characterises our model, let us explain how 
matter fields $\Phi_\alpha^{(n_\alpha)}$ can be supersymmetrically stabilised. This is possible because scalar fields can naturally acquire \acp{VEV} when demanding supersymmetry at very large energies. To see this, we must recall that heterotic orbifolds exhibit a $\U1_\text{anom}$, which induces a \ac{FI} $D$-term~\cite{Fayet:1974jb} $\xi_\text{FI}\propto \sum_\alpha q_\alpha^{\text{anom}}>0$ (the sign can always be chosen this way). Hence, the complete $D_\text{anom}$ term is
\begin{equation}
\label{eq:FIterm}
  D_\text{anom} ~=~ \xi_\text{FI} + \sum_\alpha q^\mathrm{anom}_\alpha |\Phi_\alpha^{(n_\alpha)} |^2\;.
\end{equation}
To retain supersymmetry at or close to the compactification scale, one must demand $\vev{D_\text{anom}}=0$, which leads to non-vanishing \acp{VEV}, $\vev{\Phi_\alpha^{(n_\alpha)}}$. Furthermore, it is known that also the $F$-terms can be simultaneously cancelled with $D$-flat \ac{VEV} configurations (while also $\vev{W}\sim0$)~\cite{Lebedev:2007hv,Brummer:2010fr,Kappl:2010yu}. Due to the large amount of fields solving these equations, there is enough freedom to fix by hand some of the \acp{VEV} without significantly affecting the results. However, it could be interesting to study this in detail, which will be done in an extension of this work.

This assumption has various consequences. As mentioned before, exotic matter decouples at the scale of the \acp{VEV} and some extra (gauge and flavour) symmetries are spontaneously broken. In addition, as we shall shortly see, the effective action can be simplified.
Further, in certain scenarios based on modular flavour symmetries, matter \acp{VEV} can even be instrumental to achieve realistic configurations allowing for \ac{dS} vacua~\cite{Knapp-Perez:2023nty}.

%%%%%%%%%%%%%%%%%%%%%%%%%%%%%%%%%%%%%%%%%%%%%%%%%%%%%%%%%%%%%%%%%%%%%
\subsection{Modular-Invariant Effective Action}
\label{sec:sugra}

At energies well below the compactification scale, the \ac{LEEFT} describing the dynamics of moduli and matter fields in heterotic orbifold compactifications is captured by a four-dimensional $\mathcal{N}=1$ supergravity theory. The effective action is fully specified by the K\"ahler potential $K$, superpotential $W$, and gauge kinetic functions $f_a$, all of which are subject to stringent constraints from target-space modular invariance. This discrete symmetry, inherited from the underlying two-dimensional conformal field theory, governs the transformation properties of the moduli and the matter fields, ensuring consistency of the full quantum theory.

To preserve modular invariance, the functions $K$, $W$, and $f_a$ must transform covariantly under modular transformations. In particular, the scalar potential derived from these quantities must remain invariant under the full modular group or its congruence subgroups, depending on the compactification geometry.

In addition, the modular symmetry may be anomalous at the quantum level. The mixed modular-gauge anomalies are cancelled via a \ac{GS} mechanism~\cite{Derendinger:1991hq,Ibanez:1992hc}, which induces a non-trivial modular transformation for the dilaton:
\begin{equation}\label{eq:Stransformation}
S ~\xmapsto{~~}~ S - \frac{\delta_\text{GS}}{8\pi^2}  \log(c\tau+d)\,,
\end{equation}
where $\delta_\text{GS}$ is the universal \ac{GS} coefficient and $\tau$ is the relevant K\"ahler modulus transforming under the modular group \SL{2,\Z{}}. The remaining anomaly is cancelled by threshold corrections from massive string states.

Under $\gamma\in\SL{2,\Z{}}$, the K\"ahler potential and the superpotential transform as
\begin{equation}\label{eq:KandWtransformation}
K ~\xmapsto{~\gamma~}~ K + \log|c\tau + d|^2\,,\qquad
W ~\xmapsto{~\gamma~}~ \frac{W}{(c\tau+d)}\,,
\end{equation}
so that the K\"ahler-invariant function $G = K + \ln|W|^2$
and the scalar potential remain invariant. Modular forms with definite weights $(n)$
are instrumental to construct modular-invariant quantities in the effective theory.

%%%%%%%%%%%%%%%%%%%%%%%%%%%%%%%%%%%%%%%%%%%%%%%%%%%%%%%%%%%%%%%%%%%%%
\subsubsection{The K\"ahler Potential}

The K\"ahler potential encodes the kinetic terms for both moduli and matter fields. For the untwisted K\"ahler modulus $ \tau $ and the dilaton $ S $, the K\"ahler potential including one-loop effects, takes the  form~\cite{Parameswaran:2010ec}
\begin{equation}
    \label{eq:Kahler}
     K=-\log\left[S+\overline{S} - \frac{1}{8\pi^{2}}\delta_{\text{GS}}\log\left(\I\overline{\tau}- \I \tau\right) \right] 
     -\log\left(\I\overline{\tau}- \I \tau\right)\, .
 \end{equation} 
 Here, we do not display the matter contributions,
 as they may be considered negligible in the large-volume and small matter-field \ac{VEV} limit, which we adopt here (see~\Cref{App:concrete_model} 
 for the complete expression).
Twisted matter fields $ \Phi_\alpha^{(n_\alpha)} $ localised at orbifold fixed points transform under the modular symmetry with definite modular weights $ n_\alpha $ as specified in \Cref{eq:matterp}.

%%%%%%%%%%%%%%%%%%%%%%%%%%%%%%%%%%%%%%%%%%%%%%%%%%%%%%%%%%%%%%%%%
\subsubsection{The Superpotential}

The superpotential receives two key contributions: Yukawa couplings from matter-field interactions and non-perturbative terms from gaugino condensates. Both ingredients are tightly constrained by modular symmetry.

\paragraph{Yukawa interactions.}

In heterotic orbifolds, Yukawa couplings are constrained by string selection rules~\cite{Font:1988tp,Font:1988nc,Kobayashi:2011cw} (see also~\cite{Kobayashi:2025ocp} for non-Abelian orbifolds) and modular (flavour) symmetries~\cite{Lauer:1989ax,Chun:1989se,Lauer:1990tm,Nilles:2020kgo,Baur:2024qzo}. These couplings can be computed explicitly from worldsheet \ac{CFT} correlators and typically exhibit modular weight dependence. In particular, considering twisted matter fields of the type $\Phi_\alpha^{(-\nicefrac23)}$, the leading $T'$-invariant trilinear couplings arise from
\begin{equation}
    W ~\supset~ \left(\hat Y^{(1)}_{\rep2''}(\tau)\otimes \Phi_\alpha^{(-\nicefrac23)} \otimes \Phi_\beta^{(-\nicefrac23)} \otimes \Phi_\gamma^{(-\nicefrac23)}\right)_{\rep1}\,,
\end{equation}
whose explicit expression is given in \Cref{App:concrete_model}, see \Cref{eq:Wyuk1}. Once the matter fields acquire \acp{VEV}, the effective Yukawa superpotential is given by
\begin{equation}\label{eq:Wyuk}
W_{\text{Yuk}} ~=~ -\frac{1}{\sqrt{2}}\tilde{\lambda}_1 \hat{Y}_{1}(\tau) + \tilde{\lambda}_2 \, \hat{Y}_{2}(\tau)  + \ldots\,,
\end{equation}
where $\hat{Y}_{i}(\tau)$ are modular forms defined in \Cref{eq:modularforms}, and the coefficients $\tilde{\lambda}_i$ parameterise the matter \acp{VEV}. These coefficients take the general form $\tilde\lambda_i\propto\vev{\Phi_{\alpha,p}}\vev{\Phi_{\beta,q}}\vev{\Phi_{\gamma,r}}$ up to some known coefficients and (gauge and flavour) singlet \acp{VEV}, with $\Phi_{\alpha,p}, p=1,2,3$, the components of $\Phi_\alpha^{(-\nicefrac23)}$. The factor $-1/\sqrt{2}$ 
in the first term shows the relative coupling strength between matter fields localised at the same fixed point and those localised at three different fixed points. This ratio is fixed by demanding invariance under the additional $\Delta(54)$ symmetry, which is present in these constructions~\cite{Nilles:2020kgo}. 

%%%%%%%%%%%%%%%%%%%%%%%%%%%%%%%%%%%%%%%%%%%%%%%%%%%%%%%%%%%%%%%%%
\paragraph{Non-perturbative contributions.}

Gaugino condensation in hidden sector gauge groups $G_a \subset \E8\x\E8$ generates non-perturbative superpotential terms of the form~\cite{Cvetic:1991qm}
\begin{equation}
W_{\rm gc}^{(a)} ~\sim~  \e^{- \frac{8\pi^2}{b_a} f_a(S,\tau)}\,,
\end{equation}
where $b_a$ is the one-loop beta function coefficient of $G_a$, and $f_a$ is the gauge kinetic function. At tree-level, $f_a = k_a S$, but one-loop threshold corrections introduce moduli dependence:
\begin{equation}
f_a ~=~ k_a S + \frac{1}{4\pi^2} \Delta_a(\tau)\,,
\end{equation}
with $\Delta_a(\tau)$ containing modular functions that transform appropriately to cancel anomalies and ensure covariance of the full non-perturbative term.

Phenomenologically viable models are known to frequently display multiple gaugino condensates, as shown in explicit heterotic orbifold constructions~\cite{Parameswaran:2010ec}. Hence, we consider a model with two gauge sectors condensing independently. 
We parameterise these non-perturbative effects from double gaugino condensation following~\cite{Cvetic:1991qm} as
\begin{equation}
\label{eq:Wgc}
W_{\mathrm{gc}} ~=~ \frac{\Omega_{1}(S)H_{1}(\tau)}{\eta^{2}(\tau)}+\frac{\Omega_{2}(S)H_{2}(\tau)}{\eta^{2}(\tau)}\,,  
\end{equation}
with $\eta(\tau)$ the Dedekind eta function defined in \Cref{eq:etaq} and  $H_{a}(\tau)$ are the most general modular invariant 
functions without singularities in the fundamental domain, parameterised by~\cite{eb6f8332-5e83-315e-b127-1dcd16230971, lehner1964discontinuous} 
\begin{equation}\label{eq:Hdef}
    H_a(\tau) = \left(\frac{E_4(\tau)}{\eta^8 (\tau)}\right)^{n_a}\left(\frac{E_6 (\tau)}{\eta^{12}(\tau)}\right)^{m_a}
    P(j(\tau)) ~=~ (j(\tau)-1728)^{m_a/2}j(\tau)^{n_a/3}P\left(j(\tau)\right) \, , 
\end{equation}
where $n_{a},\,m_{a}$ are some non-negative integers, $P$ is a polynomial function of the Klein invariant function $j(\tau)$ given by
\begin{equation}
    j(\tau) ~=~ 1728\,\frac{E_4(\tau)^3}{E_4(\tau)^3 - E_6(\tau)^2} \,.
\end{equation}
Here, $E_{4}(\tau)$ and $E_{6}(\tau)$ are the Eisenstein series of weight 4 and 6, respectively, defined as 
\begin{equation}
\label{eq:Eisensteins}
E_4 (\tau) ~:=~ 1+240\sum_{n=1}^\infty \frac{n^3 q^n}{1-q^n} 
\qquad\text{and}\qquad
E_6 (\tau) ~:=~ 1-504\sum_{n=1}^\infty \frac{n^5 q^n}{1-q^n}  \, .
\end{equation}
Finally, $\Omega_{a}(S)$ are functions of the dilaton given by~\cite{Parameswaran:2010ec} 
\begin{equation}\label{eq:Omega_i}
    \Omega_{a}(S) ~:=~ \frac{c_{a}}{\e}\frac{b^{0}_{a}}{96\pi^{2}}\operatorname{exp}\left[ \frac{24\pi^{2}}{b_{a}^{0}} f_{a} \right]\,,
\end{equation}
where $\e$ is the Euler number, $c_a$ are unknown constants taken for convenience here as $c_{1}=1,$ $c_{2}=8\pi^{2}e$. Further, $f_{a}$ are the 1-loop gauge kinetic functions
\begin{equation}
   f_{a} ~=~ k_{a}S-\frac{b_{a}-b^{0}_{a}}{8\pi^{2}}\log{M_{d}}
\end{equation}
in terms of the level $k_a$ of the Ka\v{c}-Moody algebra, the scale $M_d$ at which all extra (exotic) matter fields are decoupled, 
the beta-function coefficient $b_a$ that includes charged matter, and the  beta-function coefficient of the pure Yang-Mills theory.

%%%%%%%%%%%%%%%%%%%%%%%%%%%%%%%%%%%%%%%%%%%%%%%%%%%%%%%%%%%%%%%%%
\subsubsection{The scalar potential}
We now have all the ingredients to write down  the F-term of the potential for the moduli we consider. This is given by 
\begin{equation}
\label{eq:Vsugra}
    V ~=~ \e^{K}\left[ K^{A\bar{B}} D_{A} W D_{\bar{B}}\bar{ W}-3| W|^{2} \right]\,,
\end{equation}
where $A,\,B$ denote  all fields present and $D_{A} W=\partial_{A} W + \partial_{A}K\,  W$, with 
\begin{equation}
\label{eq:fullW}
  W ~=~ {W}_{\rm Yuk}+{W}_{\rm gc}\,. 
\end{equation}
Using the K\"ahler potential in  \Cref{eq:Kahler} and defining the auxiliary function $Y:=S+\overline{S} - \frac{1}{8\pi^{2}}\delta_{\text{GS}}\log\left(\I\overline{\tau}-\I \tau\right)$, 
the moduli scalar potential can be written as 
{\small
\begin{equation}\label{eq:Vexplicit}
      V = \e^{K}\left[ \left|Y W_{S}- W\right|^{2} + \frac{Y}{Y-\frac{\delta_{\text{GS}}}{8\pi^{2}}}\left| \frac{\delta_{\text{GS}}}{8\pi^{2}} \left( \frac{2W}{Y}- W_{S}  \right) 
   +\I(\I\overline{\tau}-\I \tau) W_{\tau} - W \right|^{2}-3| W|^{2} \right]\,.
\end{equation}
}

With the effective $\mathcal{N}=1$ supergravity theory fully specified and the F-term scalar potential explicitly constructed, we are now in a position to investigate the vacuum structure of the model. Given the complexity of the scalar potential and the presence of multiple moduli with nontrivial couplings, we employ numerical techniques in order to explore the landscape of solutions and identify stable vacua and other physically interesting critical points consistent with the symmetries and constraints of the compactification. 

%% file: numericalsearch.tex
\section{Mapping the modular de Sitter landscape}
\label{sec:numresults}

We now proceed to chart the modular-invariant heterotic orbifold landscape, drawing on the scalar potential defined in \Cref{sec:sugra}, \Cref{eq:Vexplicit}, and examine the distribution of unstable de Sitter saddles, stable \ac{adS} vacua, and other critical points throughout moduli space.

We assume that the matter fields acquire \acp{VEV} of the same order, taken to be $\mathcal{O}(10^{-1})$ so as to remain within the regime of validity of the K\"ahler potential calculation.  
Such \acp{VEV} can lead to the parameter values
$\tilde\lambda_{1} = 3/25000$ and 
$\tilde\lambda_{2} = 3/5000000$ in \Cref{eq:Wyuk}.  
We also set $c_1 = 1$ and $c_2 = 8\pi^2\e$, so that the coefficients in \Cref{eq:Omega_i} differ in magnitude.\footnote{These values represent a convenient choice and are not unique; however, we have explicitly verified that reasonable variations of $c_1$ and $c_2$ do not modify the qualitative features of our results.}  

The parameters $b_1$, $b_1^0$, and $\delta_{\text{GS}}$ are computed from the explicit orbifold model described in \Cref{App:concrete_model} for one \SU{3} factor, while $b_2$ and $b_2^0$ are chosen to ensure that the second condensate becomes dominant at a different order in moduli expansion.  
Since the Ka\v{c}-Moody level is $k=1$ for standard heterotic orbifold models,
we take $k_1=k_2=1$.  The decoupling scale $M_d$ must lie not far below the Planck scale;\footnote{As shown in~\cite{Lebedev:2007hv}, exotic matter can decouple a few orders of magnitude below the Planck scale.} here we fix it to be  $M_d=1/65$.  
Given these choices, the detailed structure of the critical points is controlled by the integers $m_a$, $n_a$ and by the polynomial 
$P(j(\tau))$ in the modular invariant functions \eqref{eq:Hdef}, for computational simplicity, the polynomial was fixed\footnote{Other 
choices may modify the shape of the potential, leading to other results. However, studying all the possible choices for $P(j(\tau))$ is beyond 
the scope of this work.} as $P(j(\tau))=1$ and $m_a$, $n_a$ were taken\footnote{Since $H(\tau)$ is a holomorphic 
singularity-free function in the fundamental domain, it admits a locally convergent series expansion. Hence, taking higher values 
for $n_a$ and $m_a$ yields contributions that are negligible compared to those from smaller values.} within $[0,12]$.  
All parameter values used in our analysis are summarised in \Cref{tab:parameter_values}.

\begin{table}[h]
    \centering
    \begin{tabular}{|c|c|c|c|c|c|c|c|c|c|c|c|}
    \hline
\rowcolor{gray!30}
          $\tilde{\lambda}_1$    &      $\tilde{\lambda}_2$     &
     $c_1$ & 
     $c_2$ & 
     $b_1$ & 
     $b_2$ &      $b_1^0$ &      $b_2^0$ &
     $M_d$ & 
     $k_1$ & 
     $k_2$ &      $\delta_{\text{GS}}$ \\
         \hline
         \hline
          $3/25000$     & $3/5000000$  & $1$ & $8\pi^2 \e$   & $5$  & $-26$   & $-9$  & $-12$  & $1/65$     & $1$       & $1$   & $-1 $ \\
         \hline
    \end{tabular}
    \caption{Parameter values for the numerical search.}
    \label{tab:parameter_values}
\end{table}

\begin{figure}[t!]
    \centering
    \resizebox{\textwidth}{!}
    {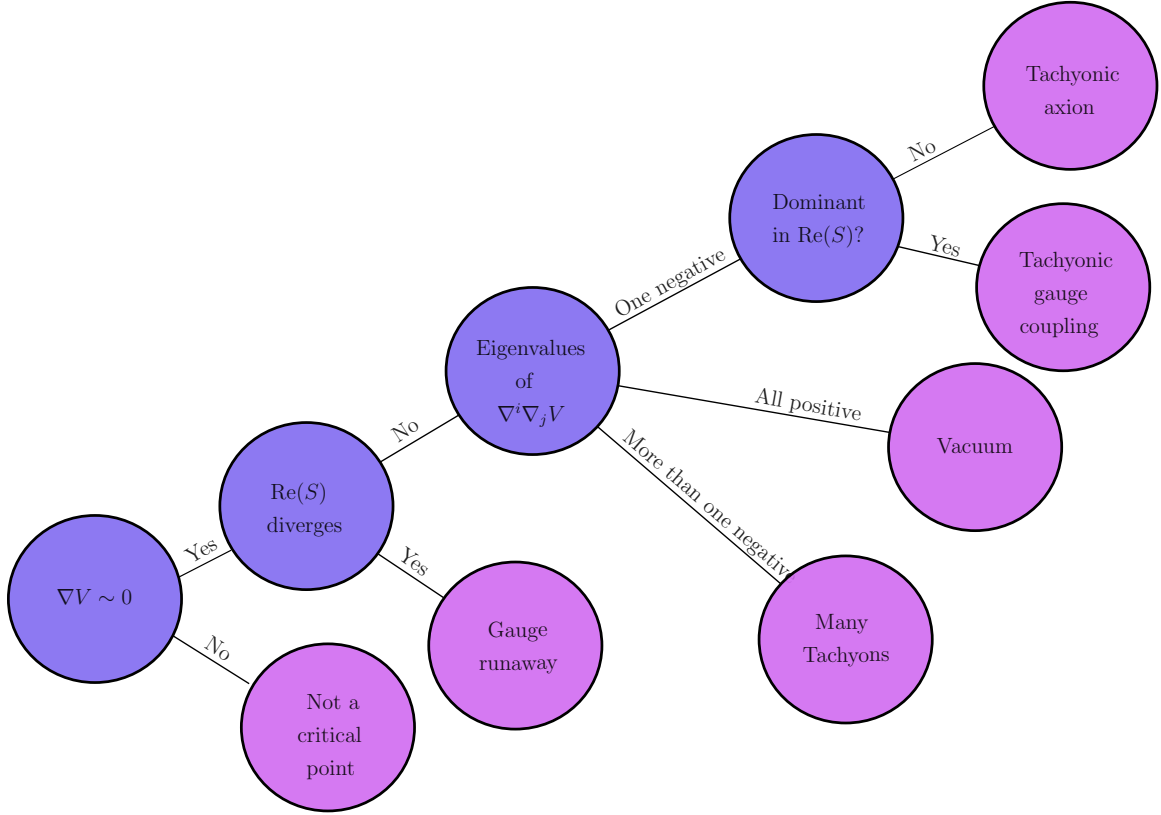}
\caption{\label{fig:Tree}
Hierarchical classification scheme applied to critical points of the scalar potential $V$, based on their properties. Initial filtering identifies candidate critical points where $\nabla V \sim 0$. Then, we discriminate the cases where $\re(S)$ diverges. The eigenvalues of the Hessian $\nabla^i\nabla_j V$  are then evaluated to assess the stability of each solution: cases with all positive eigenvalues, one negative eigenvalue, or multiple negative eigenvalues are distinguished. We then evaluate whether the dominant contribution arises from $\re(S)$. Based on the outcomes of these filters, each point is assigned to a physical class: \emph{Vacuum}, \emph{Gauge runaway}, \emph{Many tachyons}, \emph{Tachyonic axion}, \emph{Tachyonic gauge coupling}, or \emph{Not a critical point}. }    
\end{figure}

Let us first provide an overview of our procedure to classify the physics of our model.
We explore various criteria to identify phenomenologically relevant critical points of the scalar potential of \Cref{eq:Vexplicit}. 
To analyse them, we implement a hierarchical classification scheme that helps better understand their properties.  
The procedure, summarised in \Cref{fig:Tree}, begins by locating critical points, where the gradient $\nabla V$ approximately vanishes. 

In detail, our search for critical points employs both \texttt{Mathematica} and \texttt{Python} implementations.\footnote{\texttt{PyTorch} can be customised to minimise the potential; however, despite its high speed, it typically converges to minima where the dilaton runs away. Exploring improved versions is left for future work.} 
We generate $4\times 10^5$ random points uniformly distributed within the fundamental domain of the moduli space (see \Cref{fig:dominiofundamental}), scanning over the discrete parameters $n_1, n_2, m_1, m_2$. The initial values for our implemented search lie within the ranges
\begin{align}
  -0.6 &~\leq~ \re(\tau) ~\leq~ 0.6\,, \notag \hspace{2cm}
  \sqrt{1 - (\re \tau)^2} ~\leq~ \im(\tau) ~\leq~ 1.3\,, \notag \\
  0.5 &~\leq~ \re(S) ~\leq~ 2.0\,, \notag \hspace{3.7cm}
  -1 ~\leq~ \im(S) ~\leq~ 1\,.
\end{align}
As mentioned earlier, instead of focusing only on local minima, our search aims to locate as many critical points of the scalar potential in \Cref{eq:Vexplicit} as possible. For each sampled point, we compute $\nabla V$ and use \texttt{FindRoot} to solve for critical configurations.
Numerical convergence was established by requiring $|\nabla V| \lesssim 10^{-20}$ for all critical points.
If $\nabla V$ fails to vanish within tolerance (or any numerical issues arise during evaluation), the point is classified as \emph{Not a critical point}.

After identifying the critical points, they are then tested against runaway behaviour in the dilaton, signalled by  divergent values of $\re(S)$, as they indicate a negligible gauge coupling and an undesirably unstable modulus. 

To assess the stability of the critical points with a finite value of $\re(S)$, we evaluate the eigenvalues of the Hessian $\nabla^i\nabla_j V$.  
Configurations with all positive eigenvalues are labelled as (locally stable) \emph{Vacuum}, while those with one or more negative eigenvalues are classified according to the number and orientation of the tachyonic directions.  
In particular, we distinguish between solutions with a single tachyonic mode (often indicating a mild instability) and those with multiple tachyons.

The analysis is further refined by determining whether the dominant instability aligns with $\re(S)$, which dictates the gauge coupling. Configurations exhibiting this behaviour are labelled \emph{Tachyonic gauge coupling}, reflecting an instability in the gauge sector during field evolution.
Similarly, solutions where the instability combines the axions $\re(\tau)$ and $\im(S)$ are grouped as \emph{Tachyonic axion}.  
Those with multiple unstable directions are placed in the \emph{Many tachyons} category.  

\begin{figure}[t]
\centering
\includegraphics[width=1.0\linewidth]{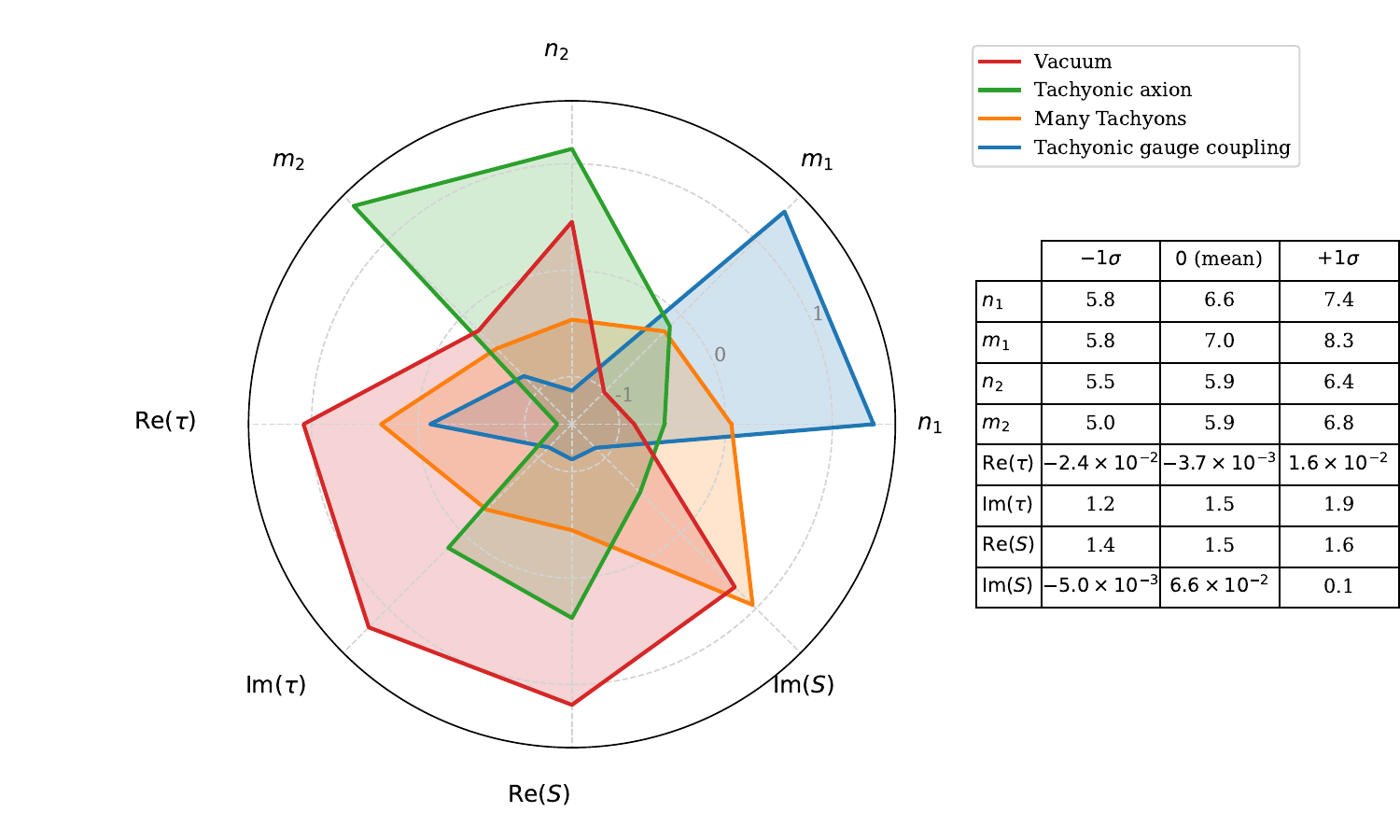}
    \caption{\label{fig:RadialZ}
    Z-score radar chart of the parameter and moduli values for the different classes of solutions. Each polygon compares how the Z-scores of the various parameters and moduli distribute, based on their mean values and $\sigma$ variance to facilitate direct comparison. The innermost (outermost) circle corresponds to the mean value minus (plus) $1\sigma$ for all variables. The table provides these statistical values for each variable in our search. The unphysical classes \emph{Not a critical point} and \emph{Gauge runaway} are omitted.}
\end{figure}

\Cref{fig:RadialZ} summarises the distribution of the values of the real components of the moduli and the integer parameters $n_a,m_a$ of \Cref{eq:Hdef}. The radial chart displays the Z-score of each variable for a better comparison. The zero Z-score, described by the second dashed circular contour from the center and marked with a 0, corresponds to the mean values of the variables. Further contours differ by a $\sigma$ deviation from the mean values. Both the mean and the deviation are given in the table of that figure. This provides a compact comparative view of the structure associated with all classes. Note that the green polygon, which indicates unstable one-dimensional saddles of $V$, includes mostly average values for $m_1$ while the parameters $n_1,n_2,m_2$ are off by about $1\sigma$ from their statistical mean in our search; furthermore, this polygon shows that e.g.\ the \ac{VEV} of the real component of the dilaton (and hence the gauge coupling) is average while $\vev{\re(\tau)}$ is well below average.

\begin{figure}[H]
\centering
\includegraphics[width=0.8\linewidth]
{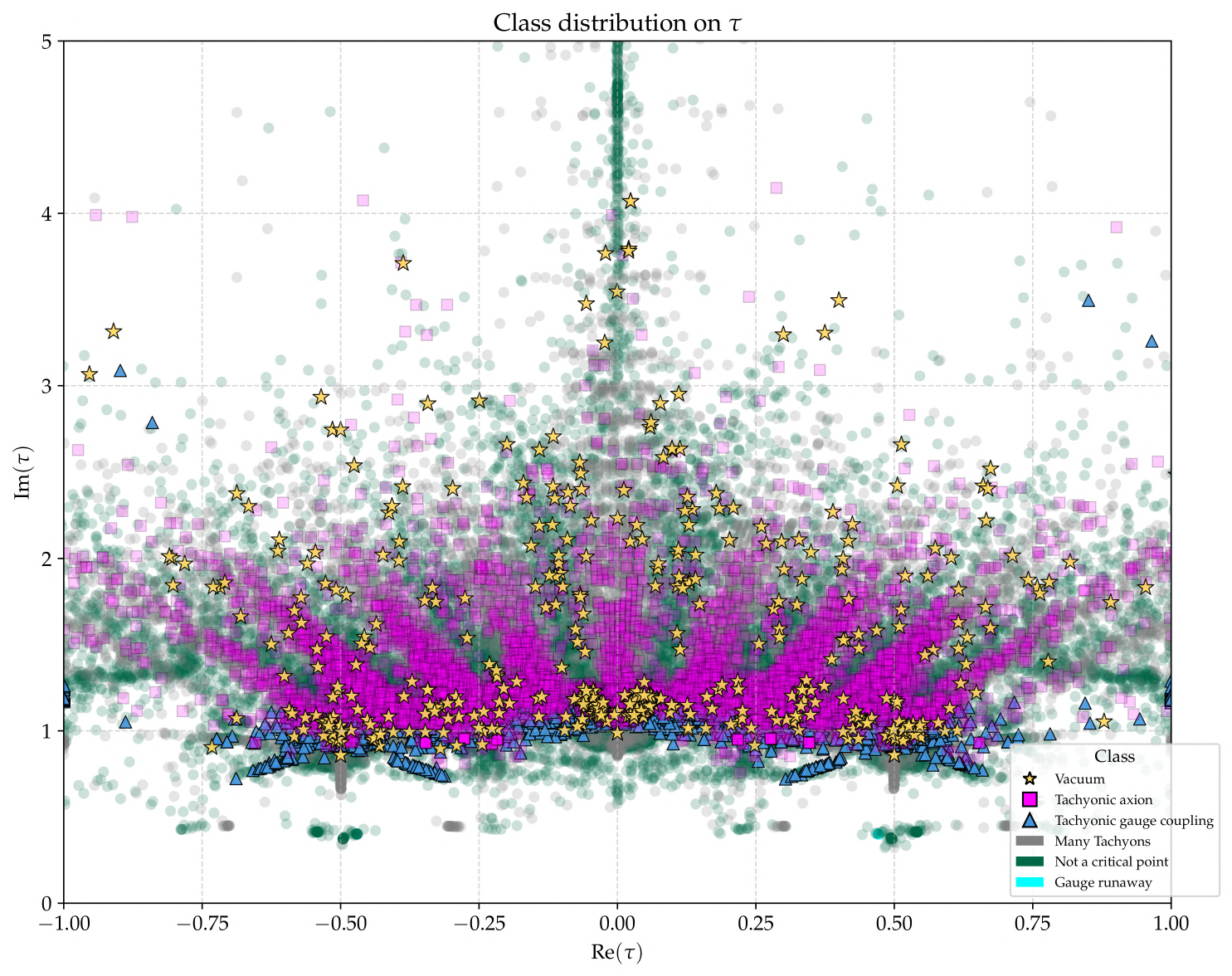}
    \caption{\label{fig:Tau}
    Distribution of the classes from \Cref{fig:Tree} in the complex $\tau$-plane. 
    Solutions corresponding to stable vacua are represented by yellow stars, whereas those identified as \textit{Tachyonic axion}  are marked with pink squares.
    \textit{Tachyonic gauge coupling}  solutions are represented
     with blue triangles, while different colours distinguish the remaining cases.
Solutions classified as \textit{Tachyonic axion} cluster in specific angular regions with a ``penacho"-like structure, while the apparent tendency of the \textit{Not a critical point} class to occupy complementary sectors is not statistically robust. 
Stable vacua occur more often near the boundaries of the fundamental domain, though they also appear sporadically throughout the region without a dominant spatial pattern.
}
\end{figure}

\begin{figure}[h]
    \centering
    \includegraphics[width=0.8\linewidth]
    {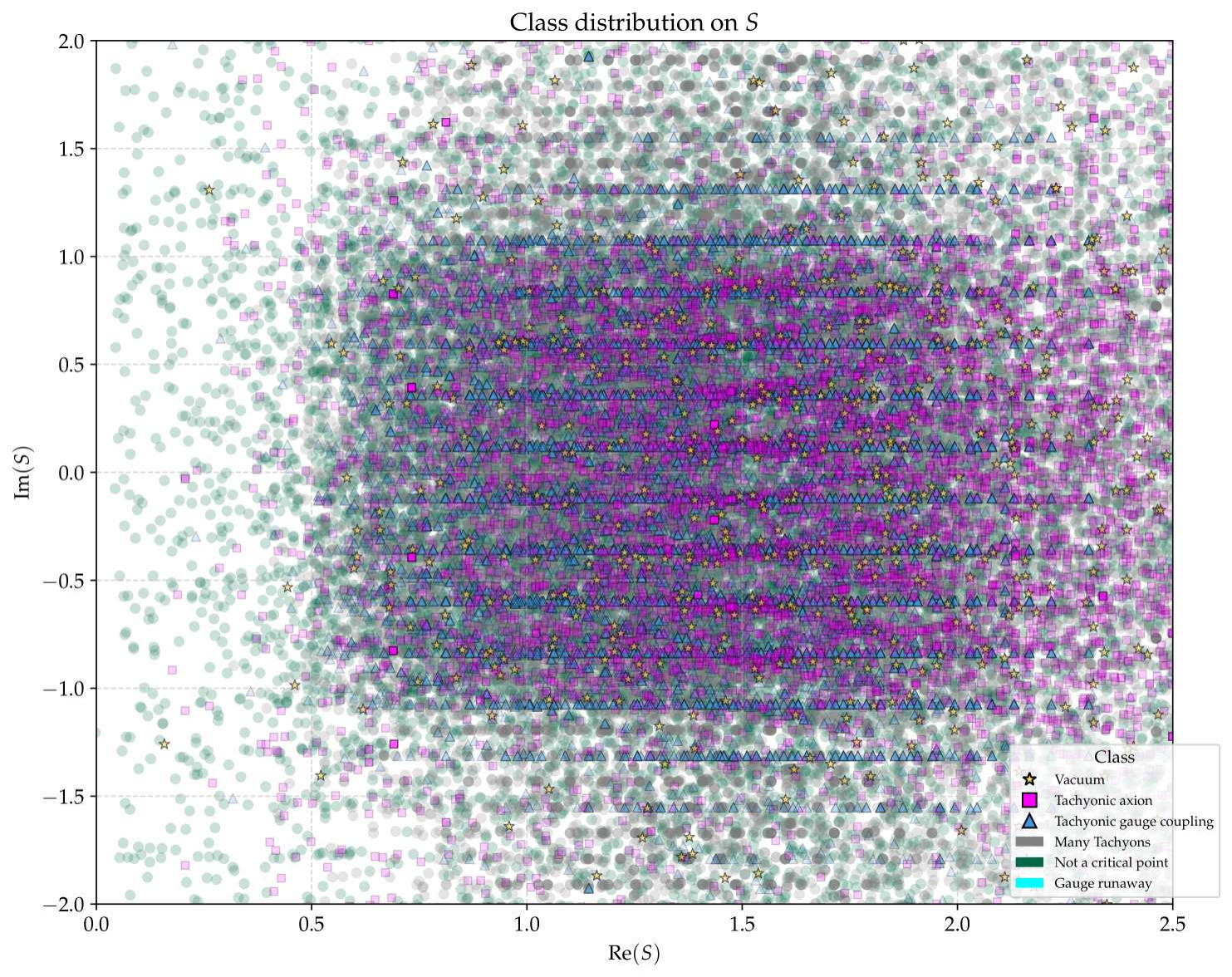}
    \caption{\label{fig:S}
    Distribution of the classes from \Cref{fig:Tree} in the dilaton plane.
    Solutions corresponding to stable vacua are represented by yellow stars, whereas those identified as \textit{Tachyonic axion}  are marked with pink squares.
    \textit{Tachyonic gauge coupling}  solutions are represented
     with blue triangles, while different colours distinguish the remaining cases.
    Solutions in the \textit{Tachyonic axion} class do not exhibit a clear spatial pattern, although they are more frequently located within the range $\im(S) \in [-1, 1]$. Solutions classified as \textit{Vacuum} are dispersed without a dominant distribution trend. Notably, \textit{Tachyonic gauge coupling} solutions tend to align periodically along horizontal bands at specific values of $\im(S)$.}
\end{figure}

Further, the resulting distributions of the various classes are shown in the complex planes of $\tau$ and $S$ in \Cref{fig:Tau,fig:S}, respectively, where  the marker shape and colour denote the class, and opacity is adjusted to highlight vacuum and tachyonic gauge coupling classes.  Concerning the possible presence of modularly equivalent points in the numerical results, we carried out an explicit verification for modular redundancies. While search seeds were drawn from the fundamental domain, we performed an \textit{a posteriori} verification to ensure the final 
dataset is free of modular redundancies. Equivalent solutions identified via the $\SL{2, \Z{}}$ mapping (including the Green-Schwarz transformation) 
constituted a negligible subset ($<0.01\%$) and were subsequently removed.
A striking feature emerges in \Cref{fig:Tau}: solutions in the \textit{Tachyonic axion} class cluster in angular regions forming a distinctive ``penacho''\footnote{The name ``penacho'' is a nod to the famed feathered headdress of Moctezuma, currently held in Vienna; it is a cultural reminder that what is displaced can still speak meaningfully to its origins.} pattern in the $\tau$-plane. The significance of this plume-like structure remains unclear, though it may suggest deeper constraints imposed by modular invariance. Further investigation is needed to determine its potential relevance.
Stable vacua appear more sparsely, typically near the boundaries of the fundamental domain.  
While \textit{Not a critical point} outcomes appear to occupy complementary regions, this effect is not statistically significant.

\Cref{fig:S} shows the distribution of the classes described in \Cref{fig:Tree} in the dilaton plane. 
The solutions cluster within the initial search range, $\re(S)\in(0.5,2.0)$, as expected. However, additional physically relevant solutions are found outside this region, favouring the weaker coupling regime ($\re(S)>2.0$) over the stronger coupling limit ($\re(S)<0.5$). Notably, the \textit{Tachyonic gauge coupling} class exhibits periodic alignment in horizontal bands, reflecting the axionic character of $\im(S)$. In contrast, the \emph{Vacuum} and \emph{Tachyonic axion} classes do not display such periodicity. A vertical band with a comparatively low density of solutions appears near $\re(S) = 1.5$, a feature that warrants further investigation. As expected, the \emph{Not a critical point} class is distributed broadly across the entire complex plane.

%%%%%%%%%%%%%%%%%%%%%%%%%%%%%%%%%%%%%%%%%%%%%%%%%%%%%%%%%%%%%%%%%%%%%%%%%%%%%%%%%%%%%%%%%%%%%%%%%%%%
\subsection{Properties of the critical points}

Until now, we have analysed the structure and distribution of the critical points of the potential according to the classification scheme of \Cref{fig:Tree}.  
Here we highlight several properties of these solutions that are relevant to the swampland programme~\cite{vanBeest:2021lhn,Grana:2021zvf} and will be important for our cosmological analysis in the next section.

%%%%%%%%%%%%%%%%%%%%%%%%%%%%%%%%%%%%%%%%%%%%%%%%%%%%%%%%%%%%%%%%
\subsubsection[Supersymmetry of adS minima]{Supersymmetry of \ac{adS} minima}\label{sec:susyadS}  
An immediate question is whether supersymmetry is preserved in the \ac{adS} minima we find.\footnote{Our model also presents \ac{adS} critical points exhibiting (one or more) tachyonic directions.}  
To address this, we compute the $F$-terms, proportional to $D_\Phi W$, along with the value of the superpotential $W$ at each minimum.  
For all minima, we find 
\begin{equation}
\langle D_\tau W \rangle \sim \mathcal{O}(10^{-19})\,, \qquad
\langle D_S {W} \rangle \sim \mathcal{O}(10^{-20})\,, \qquad
\langle {W} \rangle \sim \mathcal{O}(10^{-5})\,.
\end{equation}
Moreover, the ratio between the first term in \Cref{eq:Vsugra}, $K^{A\bar{B}} D_A {W} D_{\bar{B}} \overline{{W}}$, and the second term, $3|{W}|^2$, is always $\lesssim \mathcal{O}(10^{-23})$.  
This indicates that all \ac{adS} minima we find are supersymmetric, at least to the precision of our numerical search. 

%%%%%%%%%%%%%%%%%%%%%%%%%%%%%%%%%%%%%%%%%%%%%%%%%%%%%%%%%%%%%%%%
\subsubsection[Unstable dS]{Unstable \ac{dS}}  
Consistent with previous studies in heterotic orbifolds~\cite{Parameswaran:2010ec}, we do not find any \ac{dS} minima.  
However, we do identify a large number of \ac{dS} saddle points.  
As first noted in~\cite{Olguin-Trejo:2018zun}, such saddles can support thawing quintessence scenarios, in which scalar fields roll slowly from an initially frozen state (see~\cite{Bhattacharya:2024kxp} for cosmological constraints on the single-field saxion case, and other string-motivated hilltop constructions).  
This possibility is particularly interesting in view of recent cosmological results~\cite{DESI:2024kob,DESI:2024mwx,DESI:2025zpo,DESI:2025zgx,Lodha:2025qbg,DES:2024jxu,DES:2025bxy}, which hint at a \ac{DDE} component rather than a pure cosmological constant.  
In the next section, we focus on the class of \emph{Tachyonic axion} saddle points, which arise naturally in our modular-invariant heterotic potential and can drive multifield hilltop quintessence.

%%%%%%%%%%%%%%%%%%%%%%%%%%%%%%%%%%%%%%%%%%%%%%%%%%%%%%%%%%%%%%%%
\subsubsection{Impact of modular invariance.}  
To conclude this subsection, we investigate whether modular symmetry plays an essential role in our results.  
To this end, we explicitly break modular invariance in the potential\footnote{To this end, we modified the non-perturbative contribution to the superpotential~\eqref{eq:Wgc} as $W_{\mathrm{gc}} = \Omega_{1}(S)H_{1}(\tau)+\Omega_{2}(S)H_{2}(\tau)$.
In this way, the superpotential~\eqref{eq:fullW} does not transform as a modular form with modular weight $-1$, breaking modular invariance.} and repeat our search.  
We find that the appearance of \textit{tachyonic axion} saddle points is strongly suppressed: out of $10^5$ candidate critical points, only eight  such saddles occur without modular invariance, compared to $970$ in the modular-invariant case.  
This striking difference suggests that modular symmetry is not merely a consistency requirement, but actively shapes the vacuum structure and the presence of cosmologically relevant saddles. 

%%%%%%%%%%%%%%%%%%%%%%%%%%%%%%%%%%%%%%%%%%%%%%%%%%%%%%%%%%%%%%%%
\subsection{Tests on swampland conjectures}

The swampland program seeks to identify the criteria that an \ac{EFT} must fulfill in 
order to admit a consistent \ac{UV} completion within a theory of quantum gravity. These 
criteria are encapsulated in various swampland conjectures (see e.g.~\cite{vanBeest:2021lhn,Grana:2021zvf} for 
recent reviews), which aim to delineate the landscape of 
viable low-energy theories from the vast space of inconsistent, or swampland,  models.

Although our current setup is a string-inspired heterotic orbifold compactification, based on the important  principle of modular invariance which governs the scalar potential, it is both natural and insightful to examine the compatibility of our model with various swampland conjectures. This section is devoted to precisely that analysis.

%%%%%%%%%%%%%%%%%%%%%%%%%%%%%%%%%%%%%%%%%%%%%%%%%%%%%%%%%%%%%%%%
\subsubsection[Refined dS conjecture]{Refined \ac{dS} conjecture}

We begin with the refined \ac{dS} conjecture~\cite{Garg:2018reu,Ooguri:2018wrx} introduced in the introduction. The conjecture asserts that the scalar potential $V$ in any \ac{EFT} consistent with quantum gravity must satisfy at least one of the conditions in \Cref{eq:dSC}, reproduced here for convenience:
\begin{equation}\label{eq:dSconjecture}
\sqrt{\nabla^j V \nabla_j V } \geq c \,  V 
\qquad \text{or} \qquad 
\mathrm{min}(\nabla^i\nabla_j V) \leq -c' \, V \, ,
\end{equation}
where $c$ and $c'$ are positive constants of $\mathcal{O}(1)$, and $\mathrm{min}(\nabla^i\nabla_j V)$ denotes the smallest eigenvalue 
of the field-space covariant mass matrix $\nabla^i\nabla_j V$.

Interestingly, the unstable \ac{dS} vacua we identify do not satisfy the gradient bound, and therefore we focus on the second condition. Specifically, we evaluate the ratio 
\begin{equation}
r ~:=~ \frac{\mathrm{min}(\nabla^i\nabla_j V)}{V}
\end{equation}
to verify whether it meets the required threshold. Our analysis shows that all unstable \ac{dS} vacua in our model satisfy this bound, with values reaching as low as $r \lesssim -11$.
%-1.9$. 

These results are consistent with previous analysis on heterotic  orbifolds of \cite{Parameswaran:2010ec, Olguin-Trejo:2018zun} and \cite{Leedom:2022zdm}, where it was argued  that metastable \ac{dS} vacua can only emerge upon incorporating stringy corrections (such as Shenker-like effects~\cite{Shenker:1990uf}) into the K\"ahler potential. In their absence, no metastable \ac{dS} vacuum is viable, as it is the case in our framework.

%%%%%%%%%%%%%%%%%%%%%%%%%%%%%%%%%%%%%%%%%%%%%%%%%%%%%%%%%%%%%%%%
\subsubsection{AdS scale separation}

As discussed above, the \ac{adS} minima we obtain are all supersymmetric to the numerical precision specified earlier. Although our cosmological analysis in the next section will focus on unstable \ac{dS} saddles, the \ac{adS} vacua remain relevant: they act as the late-time end points of the rolling trajectories we find, and their properties determine the ultimate fate of the evolution. 

A particularly interesting property to examine is whether these \ac{adS} vacua exhibit \emph{scale separation}, namely a clear hierarchy between the Kaluza--Klein scale $L_{\mathrm{KK}}$ and the \ac{adS} curvature radius $L_{\mathrm{AdS}}$,
\begin{equation}\label{eq:AdSseparation}
    \frac{L_{\mathrm{KK}}}{L_{\mathrm{AdS}}} \ll 1,
\end{equation}
which would support the validity of a lower-dimensional effective description. This question has gained renewed attention in light of recent conjectures on the absence of \emph{parametric} scale separation in string theory~\cite{Gautason:2015tig,Lust:2019zwm}, although here we do not seek a parametric limit. Our analysis is instead in the spirit of studies of individual scale-separated vacua, as in e.g.~\cite{Demirtas:2021nlu,Demirtas:2021ote} (see~\cite{Coudarchet:2023mfs} for a review of the parametric conjecture and~\cite{Andriot:2025cyi, Tringas:2025uyg,Proust:2025vmv,Shiu:2022oti} for recent related work).

We define $L_{\mathrm{AdS}}$ in terms of the cosmological constant as $\Lambda=-3/L_{\mathrm{AdS}}^{2}$, and take $L_{\mathrm{KK}}$ from the standard heterotic estimate for the Kaluza--Klein mass scale,
\begin{equation}
    \Lambda_{\mathrm{KK}} ~=~ 18\sqrt{k}\,,
\end{equation}
with $k$ the Ka\v{c}--Moody algebra level~\cite{Svrcek:2006yi}.

For a statistical test, we analyse $744$ adS vacua obtained from non-equivalent models with distinct choices of the discrete parameters $(n_{1},n_{2},m_{1},m_{2})$. We find that \emph{all} vacua satisfy
\begin{equation}
    \frac{L_{\mathrm{KK}}}{L_{\mathrm{AdS}}} ~<~ 10^{-3}\,,
\end{equation}
comfortably within the bound~\eqref{eq:AdSseparation}. This indicates that the \ac{adS} vacua in our heterotic orbifold set-up admit a well-controlled 4D effective description. 

%% file: Class33.pdf_tex
%% Creator: Inkscape 1.3.2 (1:1.3.2+202311252150+091e20ef0f), www.inkscape.org
%% Accompanies image file 'Class33.pdf' (pdf, eps, ps)
\begingroup%
  \makeatletter%
  \providecommand\color[2][]{%
    \errmessage{(Inkscape) Color is used for the text in Inkscape, but the package 'color.sty' is not loaded}%
    \renewcommand\color[2][]{}%
  }%
  \providecommand\transparent[1]{%
    \errmessage{(Inkscape) Transparency is used (non-zero) for the text in Inkscape, but the package 'transparent.sty' is not loaded}%
    \renewcommand\transparent[1]{}%
  }%
  \providecommand\rotatebox[2]{#2}%
  \newcommand*\fsize{\dimexpr\f@size pt\relax}%
  \newcommand*\lineheight[1]{\fontsize{\fsize}{#1\fsize}\selectfont}%
  \ifx\svgwidth\undefined%
    \setlength{\unitlength}{662.59903074bp}%
    \ifx\svgscale\undefined%
      \relax%
    \else%
      \setlength{\unitlength}{\unitlength * \real{\svgscale}}%
    \fi%
  \else%
    \setlength{\unitlength}{\svgwidth}%
  \fi%
  \global\let\svgwidth\undefined%
  \global\let\svgscale\undefined%
  \makeatother%
  \begin{picture}(1,0.66668229)%
    \lineheight{1}%
    \setlength\tabcolsep{0pt}%
    \put(0,0){\includegraphics[width=\unitlength,page=1]{Class33.pdf}}%
    \put(0.83110693,0.44781581){\color[rgb]{0,0,0}\transparent{0.88}\rotatebox{0.417695}{\makebox(0,0)[lt]{\lineheight{1.25}\smash{\begin{tabular}[t]{l}Tachyonic \\\hspace{4pt}   gauge \\ coupling\end{tabular}}}}}%
    \put(0,0){\includegraphics[width=\unitlength,page=2]{Class33.pdf}}%
    \put(0.21345577,0.2570441){\color[rgb]{0,0,0}\transparent{0.88}\makebox(0,0)[lt]{\lineheight{1.25}\smash{\begin{tabular}[t]{l}\hspace{2pt}$\operatorname{Re}(S)$\\diverges\end{tabular}}}}%
    \put(0.15119276,0.20408514){\color[rgb]{0,0,0}\transparent{0.88}\rotatebox{31.049975}{\makebox(0,0)[lt]{\lineheight{1.25}\smash{\begin{tabular}[t]{l}Yes\end{tabular}}}}}%
    \put(0.83642855,0.60048807){\color[rgb]{0,0,0}\transparent{0.88}\rotatebox{0.41769515}{\makebox(0,0)[lt]{\lineheight{1.25}\smash{\begin{tabular}[t]{l}Tachyonic \\\hspace{6pt}    axion\end{tabular}}}}}%
    \put(0.61342174,0.33709327){\color[rgb]{0,0,0}\transparent{0.88}\rotatebox{-10.083258}{\makebox(0,0)[lt]{\lineheight{1.25}\smash{\begin{tabular}[t]{l}All positive\end{tabular}}}}}%
    \put(0.75562208,0.46321648){\color[rgb]{0,0,0}\transparent{0.88}\rotatebox{-14.547498}{\makebox(0,0)[lt]{\lineheight{1.25}\smash{\begin{tabular}[t]{l}Yes\end{tabular}}}}}%
    \put(0.7440107,0.53435491){\color[rgb]{0,0,0}\transparent{0.88}\rotatebox{26.115854}{\makebox(0,0)[lt]{\lineheight{1.25}\smash{\begin{tabular}[t]{l}No\end{tabular}}}}}%
    \put(0.50279604,0.40629951){\color[rgb]{0,0,0}\transparent{0.88}\rotatebox{29.085878}{\makebox(0,0)[lt]{\lineheight{1.25}\smash{\begin{tabular}[t]{l}One negative\end{tabular}}}}}%
    \put(0.32035327,0.20918395){\color[rgb]{0,0,0}\transparent{0.88}\rotatebox{-35.638651}{\makebox(0,0)[lt]{\lineheight{1.25}\smash{\begin{tabular}[t]{l}Yes\end{tabular}}}}}%
    \put(0.32111683,0.30315997){\color[rgb]{0,0,0}\transparent{0.88}\rotatebox{33.533261}{\makebox(0,0)[lt]{\lineheight{1.25}\smash{\begin{tabular}[t]{l}No\end{tabular}}}}}%
    \put(0.16140238,0.13593308){\color[rgb]{0,0,0}\transparent{0.88}\rotatebox{-33.242177}{\makebox(0,0)[lt]{\lineheight{1.25}\smash{\begin{tabular}[t]{l}No\end{tabular}}}}}%
    \put(0.23839184,0.084492){\color[rgb]{0,0,0}\transparent{0.88}\makebox(0,0)[lt]{\lineheight{1.25}\smash{\begin{tabular}[t]{l} \hspace{1pt} Not a \\ critical \\\hspace{1pt}  point\end{tabular}}}}%
    \put(0.50487397,0.30703531){\color[rgb]{0,0,0}\transparent{0.88}\rotatebox{-40.809442}{\makebox(0,0)[lt]{\lineheight{1.25}\smash{\begin{tabular}[t]{l}More than one negative\\\end{tabular}}}}}%
    \put(0.0407136,0.17084802){\color[rgb]{0,0,0}\transparent{0.88}\makebox(0,0)[lt]{\lineheight{1.25}\smash{\begin{tabular}[t]{l}$\nabla V \sim 0$\end{tabular}}}}%
    \put(0,0){\includegraphics[width=\unitlength,page=3]{Class33.pdf}}%
    \put(0.38528117,0.37598529){\color[rgb]{0,0,0}\transparent{0.88}\makebox(0,0)[lt]{\lineheight{1.25}\smash{\begin{tabular}[t]{l}Eigenvalues \\ \hspace{.75cm}of\\ \hspace{0.4cm}$\nabla^i\nabla_j V$\end{tabular}}}}%
    \put(0.62840006,0.49565272){\color[rgb]{0,0,0}\transparent{0.88}\makebox(0,0)[lt]{\lineheight{1.25}\smash{\begin{tabular}[t]{l}Dominant \\in $\operatorname{Re}(S)$?\end{tabular}}}}%
    \put(0,0){\includegraphics[width=\unitlength,page=4]{Class33.pdf}}%
    \put(0.38736264,0.14411579){\color[rgb]{0,0,0}\transparent{0.88}\rotatebox{0.417695}{\makebox(0,0)[lt]{\lineheight{1.25}\smash{\begin{tabular}[t]{l}\hspace{1pt} Gauge \\runaway\end{tabular}}}}}%
    \put(0,0){\includegraphics[width=\unitlength,page=5]{Class33.pdf}}%
    \put(0.65323865,0.1506503){\color[rgb]{0,0,0}\transparent{0.88}\rotatebox{0.417695}{\makebox(0,0)[lt]{\lineheight{1.25}\smash{\begin{tabular}[t]{l}\hspace{3pt}   Many \\Tachyons \end{tabular}}}}}%
    \put(0,0){\includegraphics[width=\unitlength,page=6]{Class33.pdf}}%
    \put(0.7633081,0.29493228){\color[rgb]{0,0,0}\transparent{0.88}\rotatebox{0.417695}{\makebox(0,0)[lt]{\lineheight{1.25}\smash{\begin{tabular}[t]{l}Vacuum\end{tabular}}}}}%
  \end{picture}%
\endgroup%

%% file: results.tex
\section{Multifield quintessence from modular dS saddles}
\label{sec:quintessence}

We have seen that the modular invariant potential~\eqref{eq:Vexplicit} exhibits a rich landscape of \ac{dS} saddle points, all consistent with the refined de Sitter swampland conjecture~\eqref{eq:dSC}. At the same time, recent cosmological observations hint at an evolving dark energy equation of state. Within quintessence scenarios, the data appear to favour \emph{thawing} models, where the equation of state $w_\varphi$ begins close to $-1$ and gradually increases. Hilltop quintessence~\cite{Dutta:2008qn} provides a well-studied realisation of this class and has been explored in supergravity and string-motivated settings~\cite{Olguin-Trejo:2018zun}. More recently, single-field axion and saxion hilltops have been directly confronted with the DESI and DES data in~\cite{Bhattacharya:2024kxp}.

Motivated by this, we now investigate cosmological evolution in the modular invariant heterotic potentials introduced above. In particular, we study the dynamics near a \ac{dS} saddle point in the full four-field system of the complex dilaton $S$ and 
K\"ahler modulus $\tau$. This extends earlier single-field hilltop constructions to a setting where multiple moduli can roll simultaneously, as is generically expected in string compactifications.

Concretely, we focus on a representative saddle point with a single tachyonic direction dominated by the dilaton axion:
\begin{equation}
\im(\psi^S) ~\sim~ 0.9\,\im(S) - 0.4\,\re(\tau)\,,
\end{equation}
belonging to the \emph{Tachyonic axion} branch of our classification in \Cref{sec:numresults} (see \Cref{fig:Tree}). Starting from this configuration, we evolve the fields along their cosmological trajectory as they roll towards a nearby \ac{adS} minimum. The parameter values and vacuum expectation values at the saddle are listed in~\Cref{table:parameters}.

The dynamics are governed by the coupled Einstein–scalar system:
\begin{subequations}\label{subeq:eoms}
\begin{align}
3\left(\frac{\dot{a}}{a}\right)^2 &~=~ \frac{1}{2} g_{ij}\,\dot{\varphi}^i \dot{\varphi}^j + V
+ 3H_0^2 \Omega_{M,0}\, a^{-3} + 3H_0^2 \Omega_{r,0}\, a^{-4}\,, \label{eq:HubbleP} \\
0 &~=~ \ddot{\varphi}^i + 3 \frac{\dot{a}}{a} \dot{\varphi}^i + \Gamma^i_{\; jk} \dot{\varphi}^j \dot{\varphi}^k 
+ g^{ij}\,\partial_j V \,, \label{eq:motioneq}
\end{align}
\end{subequations}
where we assume a flat \ac{FLRW} metric with scale factor $a(t)$ and we include matter and radiation, whose contributions are encoded in the density parameters today,
\begin{equation}
\Omega_{i,0} ~:=~ \frac{\rho_{i,0}}{3H_0^2}\,, 
\qquad\text{with}\qquad
\rho_i ~\propto~ a^{-3(1+w_i)}
\end{equation}
and $w_i$ the equation of state of each component. The scalar fields are denoted together as
\begin{equation}
\label{eq:phivector}
\varphi ~=~ \left(\varphi^i\right) ~=~ \big( \re(\tau),\, \im(\tau),\, \re(S),\, \im(S)\big)\,,
\end{equation}
with field-space metric
\begin{equation}
\frac{\partial^2 K}{\partial \Psi^I \partial \overline{\Psi}^J}\,\partial \Psi^I 
\partial \overline{\Psi}^J ~=~ \tfrac{1}{2}\, g_{ij}\, \partial \varphi^i \partial \varphi^j\,,
\end{equation}
for $\Psi^I=\{\tau,S\}$. The scalar potential $V$ is given in~\Cref{eq:Vexplicit}.
We stress that in our analysis we incorporate one-loop corrections, which appear both in the scalar potential and in the K\"ahler metric. These corrections are crucial for a consistent heterotic orbifold description and imply that the field-space metric is no longer diagonal. In fact, at the saddle point under consideration, the corrected metric takes the form
\footnotesize
\begin{equation}
g_{\mathrm{1-loop}} ~=~
\begin{pmatrix}
0.298 & 0 & 0 & -0.00125 \\
0 & 0.298 & 0.00125 & 0 \\
0 & 0.00125 & 0.256 & 0 \\
-0.00125 & 0 & 0 & 0.256
\end{pmatrix}\,,
\end{equation}
\normalsize
to be contrasted with the tree-level result, which is purely diagonal.

To match the present-day universe, we impose that the cosmological evolution reproduces the observed density fractions and the effective equation of state today. Although these quantities are somewhat model-dependent, significant deviations from the $\Lambda$CDM values are not expected. As fiducial values, we adopt those reported by Planck~\cite{Planck:2018vyg}:
\begin{equation}\label{eq:Omega0values}
\Omega_{m,0} ~=~ 0.3111\,, \qquad 
\Omega_{r,0} ~=~ 0.0001\,, \qquad 
\Omega_{\varphi,0} ~=~ 0.6889%0.6883
\,,
\end{equation}
as well as $H_0=5.927 \x 10^{-61}% M_\mathrm{Pl}
$.

\begin{table}[t]
\centering
\renewcommand{\arraystretch}{1.3} % spacing
\setlength{\tabcolsep}{10pt}      % column spacing
\resizebox{\textwidth}{!}{
\begin{tabular}{|c | c |c|}
\hline
\rowcolor{gray!30}
\multicolumn{3}{|c|}{\textbf{Parameters}} \\
\hline
$\tilde\lambda_{1}=3/25000$, $\tilde\lambda_{2}=3/5000000$  
& $m_{1}=12, \, n_{1}=0, \, m_{2}=10, \, n_{2}=3$ 
& $b_1^0=-9, \, b_2^0=-12, \, A = 6.70\x 10^{-112}$ \\ 
\hline\hline

\rowcolor{gray!30}
\multicolumn{3}{|c|}{\textbf{Moduli \acp{VEV}}} \\
\hline
$\Psi^I$ & Re$\langle \Psi^I \rangle$ & Im$\langle \Psi^I \rangle$\\
\hline
$\tau$ & $-0.004063$ & $1.297$ \\ 
$S$    & $1.390$ & $0.1272$ \\
\hline\hline

\rowcolor{gray!30}
\multicolumn{3}{|c|}{\textbf{Mass eigenstates}} \\
\hline
$\psi^I \sim \Psi^I$ & $m^2_{\re(\psi^I)}$ & $m^2_{\im(\psi^I)}$ \\ 
\hline
$\psi^\tau$  & $3.190\x 10^{-116}$ &  $3.292\x 10^{-116}$\\ 
$\psi^S$    & $2.830\x 10^{-119}$ & $-3.574\x 10^{-119}$ \\ 
\hline
\rowcolor{gray!10}
\multicolumn{3}{|c|}{\textit{Tachyon:} $\im(\psi^S)\sim -0.4\, \re(\tau) + 0.9\, \im(S)$}\\
\hline
\end{tabular}}
\caption{\label{table:parameters}
Axionic saddle point of the potential~\eqref{eq:Vexplicit}. The values of $m^2$ correspond to
the eigenvalues of the mass matrix $\nabla^i\nabla_j V$.
The mass eigenstate $\psi^I$
has a dominant contribution of the modulus $\Psi^I$.
The unstable direction is mostly governed by the axionic fields 
$\re(\tau)$  and $\im(S)$. All \acp{VEV} and masses are given in reduced Planck units with $M_\mathrm{Pl}=1$.}
\end{table}

\begin{figure}[t]
  \centering
          \includegraphics[width=0.63\textwidth]{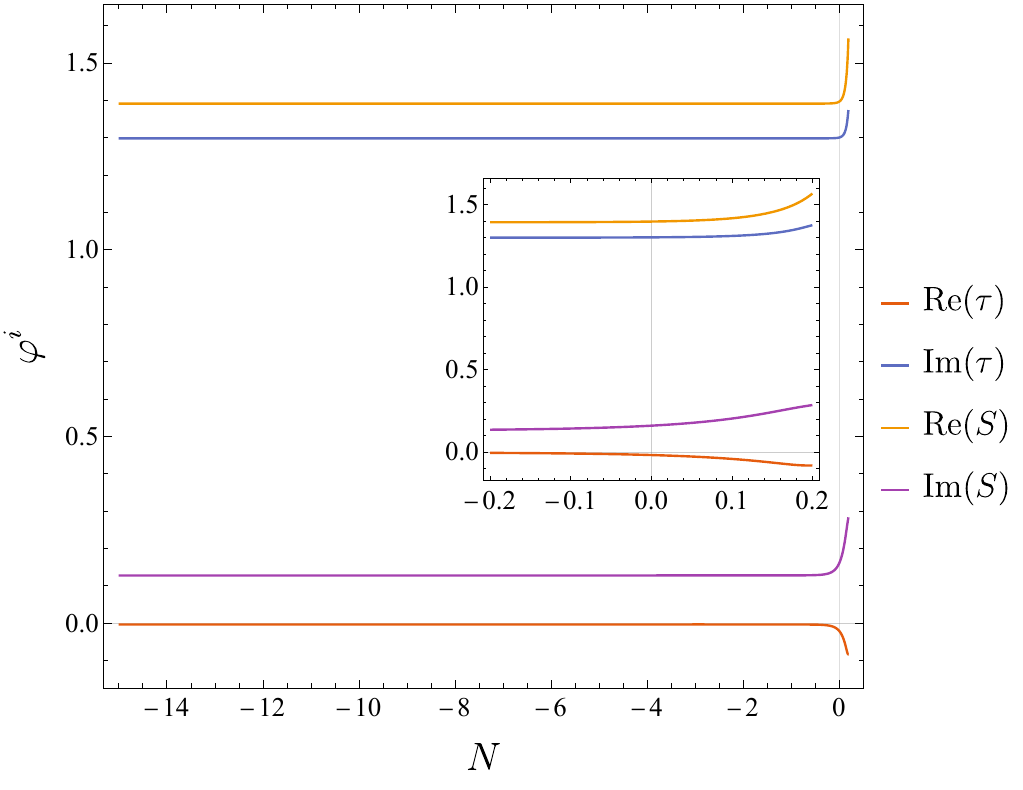}
    \caption{\label{fig:FieldsEv}
    Evolution of the fields $\varphi^i$ as function
    of the number of e-folds $N$. These curves represent the numerical solutions to the equations of motion \eqref{subeq:eoms}, using the parameters listed in \Cref{table:parameters}. The initial conditions are slightly displaced from the saddle point by $\delta\varphi=(-0.00025,\, 0.0001,\, 0.0001,\, -0.0001)$. Here, $N=-15$ corresponds to some time between BBN and matter-radiation-equality, while $N=0$
    represents the present time.}
\end{figure}

\vspace{1cm}
We solve the equations of motion~\eqref{subeq:eoms} numerically, using the number of e-folds 
$N=\ln a$ as the evolution variable instead of cosmic time. 
As is well known in hilltop-like quintessence, if the fields start exactly at the saddle point, 
they remain there throughout the entire cosmological history, with the potential energy 
contributing as a cosmological constant. For \ac{DDE} to arise, however, the fields must be 
slightly displaced from the saddle, after which they begin to roll down their potential.

Accordingly, we start the fields with a small displacement from the saddle point,
\begin{equation}\label{eq:displacement}
    \varphi_{\rm init}^i ~=~ \vev{\varphi_\mathrm{saddle}^i}+\delta\varphi^i\, , 
\end{equation}
with $\delta\varphi=(-0.00025,\, 0.0001,\, 0.0001,\, -0.0001)$, where $\varphi$ is defined in \Cref{eq:phivector}.
The initial conditions must be extremely close to the saddle in order to prevent the energy density from becoming negative before the present epoch, which would otherwise trigger a rapid recollapse. 
In fact, ensuring that the quintessence equation of state satisfies $w_\varphi<-1/3$ to the present day requires $\delta\varphi^i\lesssim 10^{-3}$. In addition, the potential value at the saddle point must be fine-tuned, as is generic in quintessence scenarios. Here we rescale the potential by an overall factor,
\begin{equation}
    \widetilde{V} ~=~ A\, V\, , \qquad 
    A ~=~ 6.7\x10^{-112}\,,
\end{equation}
so as to reproduce the observed small value of dark energy today. This rescaling may be related to 
the \acp{VEV} of matter fields and to the polynomials $P(j(\tau))$ associated with gaugino 
condensates in \Cref{eq:Hdef}.

The evolution of the four real scalar fields is displayed in \Cref{fig:FieldsEv}. We take $N=-15$, which corresponds to an epoch between \ac{BBN} and matter–radiation equality (the latter occurring at $N\simeq -8.1$), while matter–dark energy equality happens around $N\simeq -0.26$, and $N=0$ denotes the present time. The integration is continued until $N\simeq 0.2$ (about $3.0\x10^9$ years into the future\footnote{One e-fold today corresponds to roughly $14.5$\,Gyr, hence $0.2$ e-folds correspond to $\sim 3$ Gyr.}) to capture the full trajectory of the fields after they drive cosmic acceleration.

As shown, all four fields remain frozen for most of the cosmological history and only begin to roll very recently, thereby driving the present accelerated expansion. 
The K\"ahler modulus is stabilised at $\tau\simeq \I$ up to the present epoch, 
a point that has been extensively studied in flavour phenomenology, where it leads to realistic 
fermion mass hierarchies~\cite{Novichkov:2021evw, Feruglio:2021dte}. 
Meanwhile, the real part of the dilaton is stabilised at $\re(S)\simeq 1.390$, 
yielding the universal 4-dimensional gauge coupling
\begin{equation}
  g_4^2 ~=~ \frac{1}{\re\,\vev{S}} ~\simeq~ 
  % 0.848
 0.719 \, .    
\end{equation}
This value is of order unity, consistent with the expectation that the effective gauge couplings 
in heterotic orbifolds should be moderately strong, lying near the edge of the perturbative regime where control is still maintained.
Finally, we note that in the far future the fields evolve towards a nearby \ac{adS} minimum, 
as will be discussed in \Cref{ssec:AdSVacuum}.

\begin{figure}[H]
  \centering
    \subfloat{%
    \includegraphics[width=0.47\textwidth]{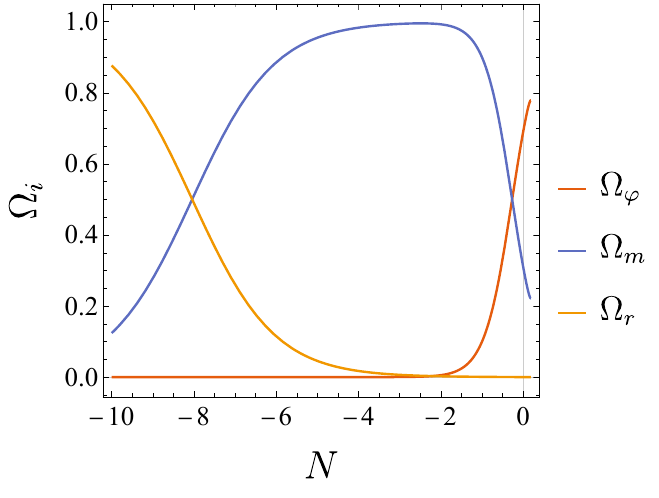}
    }
    \hspace{0.25cm}
    \subfloat{%
    \includegraphics[width=0.40\textwidth]{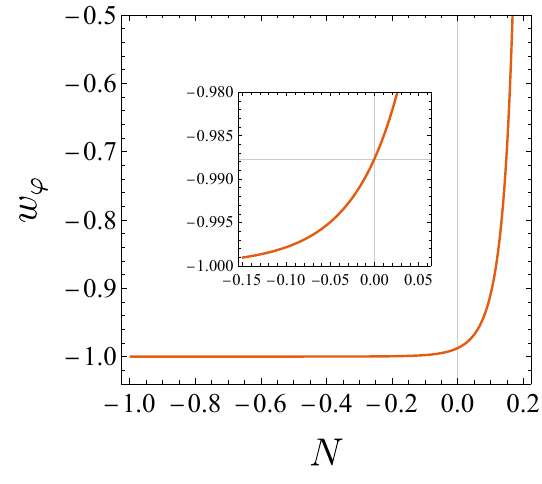}
    }  
    \caption{\label{fig:Observables}
    Evolution of the cosmological observables as functions of the number of e-folds $N$. 
    Their current values are $\Omega^\text{mod}_{\varphi,0}~\simeq~ 0.6883,\Omega^\text{mod}_{m,0}~\simeq~ 0.3116,\Omega^\text{mod}_{r,0}~\simeq~0.0001$ and $w^\text{mod}_{\varphi,0}~\simeq~ -0.9877$.
    }
\end{figure}

\Cref{fig:Observables} shows the evolution of the quintessence equation of state $w_\varphi$ and the density fractions $\Omega_i$. 
As expected, the fields roll extremely slowly until very recently, keeping 
$w\simeq -1$ for most of cosmological history. At present, we find in our model
\begin{equation*}
w^\text{mod}_{\varphi,0}~\simeq~ -0.9877
\qquad \text{and}\qquad
\Omega^\text{mod}_{\varphi,0}~\simeq~ 0.6883\, ,
\end{equation*}
in excellent agreement with observations. 

%%%%%%%%%%%%%%%%%%%%%%%%%%%%%%%%%%%%%%%%%%%%%%%%%%%%%%%%%%%%%%%%%%%%%%%%%%%%%%%%%%%
\subsubsection*{Field displacement and the distance conjecture}

We can further compute the total geodesic displacement of the scalar fields from $N=-15$ up to 
today ($N=0$), obtaining $\Delta \varphi\simeq 0.16 $. This value lies safely within the bound 
imposed by the so-called \emph{distance conjecture}. This conjecture~\cite{Ooguri:2006in} 
states that in any \ac{EFT} consistent with quantum gravity, the field space displacement must satisfy
\begin{equation}
    \Delta \varphi ~\lesssim~ \tilde{c}  \, ,
\end{equation}
with $\tilde{c}\sim \mathcal{O}(1)$. Exceeding this bound would signal the appearance of an infinite 
tower of states becoming exponentially light,
\begin{equation}
    m ~\sim~ \e^{-\alpha \Delta \varphi}\, , \qquad
    \alpha ~\sim~ \mathcal{O}(1)\, ,
\end{equation}
which invalidates the \ac{LEEFT} description. This reflects the expectation that large excursions in 
field space trigger the emergence of new degrees of freedom, signalling the breakdown of the 
effective description. 

It should be emphasised that our computation does not include the very early universe, e.g.\ possible epochs immediately after inflation. For instance, a kination phase is possible in scalar 
models, during which $\rho\propto a^{-6}$ and the field displacement could be substantially larger. 
However, such epochs are highly model dependent and are not guaranteed to occur or last for long in the present setup.

%%%%%%%%%%%%%%%%%%%%%%%%%%%%%%%%%%%%%%%%%%%%%%%%%%%%%%%%%%%%%%%%%%%%%%%%%%%%%%%%%%%
\subsection[\texorpdfstring{Cosmological constraints on $w_\varphi$}{Cosmological constraints on w\_varphi}]{\boldmath Cosmological constraints on $w_\varphi$ \unboldmath}\label{ssec:compareDESI}

Although we do not perform a full likelihood analysis against cosmological data, we can nevertheless confront our background evolution with observational constraints, in particular on the dark energy equation of state. 
We focus on the most recent DESI-DR2 results, which report constraints on $w_{\rm DE}$ based on the commonly used \ac{CPL} parametrisation~\cite{Chevallier:2000qy, Linder:2002et}, where the dark energy equation of state is expressed as
\begin{equation}
w_{\rm DE}(a) ~=~ w_0 + (1 - a)\, w_a \, ,
\end{equation}
namely a first-order Taylor expansion of $w_{\rm DE}$ in the scale factor $a$, with linear leading behaviour. 
Considering the combined DESI+CMB+Union3 datasets~\cite{DESI:2025zgx}, the reported constraints are
\begin{equation}\label{eq:w0wadesi}
    w_0 ~=~-0.667\pm 0.088 
    \qquad\text{and}\qquad  
    w_a ~=~ -1.09^{+0.31}_{-0.27} \, .
\end{equation}
While our saddle quintessence model is not expected to be captured by the CPL ansatz, it is still instructive to compare. 
We find that by slightly adjusting the initial displacement from the saddle point, one can obtain values of $w_0$ closer to those in~\eqref{eq:w0wadesi}. 
For instance, displacing the fields as
\begin{equation}
\delta\varphi ~=~ 
  (0.000641,\, 0.0001,\, 0.0001,\, 0.000185)\,,
\end{equation}
yields a present-day value $w_0 = -0.838$. Larger values of $w_0$ closer to $-1$, or significantly higher values, can be obtained similarly with suitable initial conditions. 
This sensitivity is in fact a well-known feature of hilltop quintessence: the evolution of $w_\varphi$ is highly dependent on both the displacement from the maximum and the local curvature of the potential~\cite{Dutta:2008qn}. 
We find that for displacements $\delta\varphi^i_{\rm max} \gtrsim 10^{-3}$, the fields are too far from the saddle and roll too fast to remain compatible with acceleration today. 

\begin{figure}[t]
  \centering
   \includegraphics[width=0.55\textwidth]{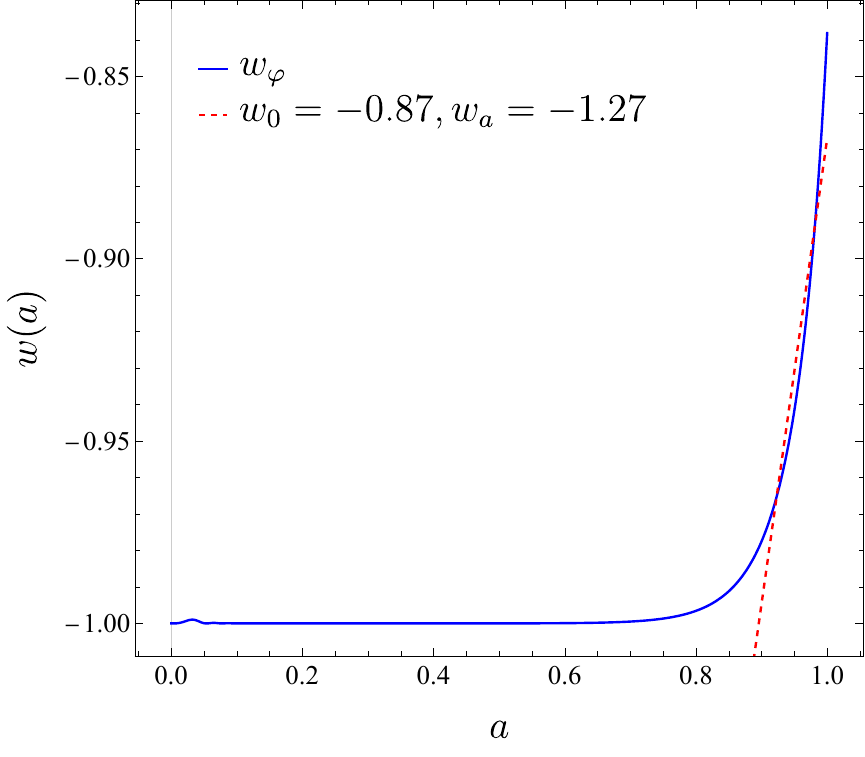}
    \caption{\label{fig:CPLpar}
    Equation of state parameter $w$ as a function of the scale factor $a$ for our quintessence model $\varphi$, compared with the \ac{CPL} parametrization of DESI~\cite{DESI:2025zgx}. 
    The linear fit was performed between $a = 0.9$  and $a = 1.0$. The red dashed line yields a chi-squared error of $\chi^2 \simeq 5.6$, where we have compared with DESI+CMB+Union3 data.
    }
\end{figure}

In \Cref{fig:CPLpar} we show the comparison between the CPL parametrisation and our saddle quintessence evolution. The CPL parameters were obtained by fitting the model’s approximately linear behaviour in the interval $a = 0.9$ to $a = 1$. 
When compared with the DESI+CMB+Union3 results, our fit yields a best-fit deviation corresponding to $\chi^2 \simeq 5.6$. 

A more accurate parametrisation for single-field hilltop quintessence was obtained  by Dutta–Scherrer~\cite{Dutta:2008qn} and extended to general thawing models by Chiba~\cite{Chiba:2009sj} (DSCh). This parametrisation was obtained considering an expansion around the maximum (initial value of $\phi$ in \cite{Chiba:2009sj}) of the potential
$V(\phi) = V(\phi_\mathrm{max}) + V''(\phi_\mathrm{max})(\phi - \phi_\mathrm{max})^2$.
However, since our scenario involves four coupled scalar fields, this assumption cannot be made and thus a large discrepancy will arise. Developing an accurate parametrisation for multifield saddle quintessence thus remains an interesting task for future work.

%%%%%%%%%%%%%%%%%%%%%%%%%%%%%%%%%%%%%%%%%%%%%%%%%%%%%%%%%%%%%%%%%%%%%%%%%%%%%%%%%%%
\subsection{The fate of dark energy: descent into adS}
\label{ssec:AdSVacuum}

Thus far we have focused on the multifield dynamics relevant for today's universe. 
It is, however, interesting to let the fields evolve further in the future to determine the fate of today's accelerated expansion in the class of models we study. 
In particular, in the saxion supergravity hilltop of~\cite{Olguin-Trejo:2018zun,Bhattacharya:2024kxp}, the late-time dynamics drive the system towards a runaway at large field values. 
Since the potential in that case is unbounded from below at small saxion values, an additional fine-tuning was required to ensure that the field started slightly displaced from the hilltop, rolling towards larger values. 
In contrast, in the present setup we can displace the fields from the saddle either towards smaller or larger values of the four scalars. This allows us to track the full evolution of the system and determine, without additional tuning, the eventual fate of dark energy in our model. 

\begin{figure}[H]
  \centering
    \subfloat{%
      \includegraphics[width=0.45\textwidth]{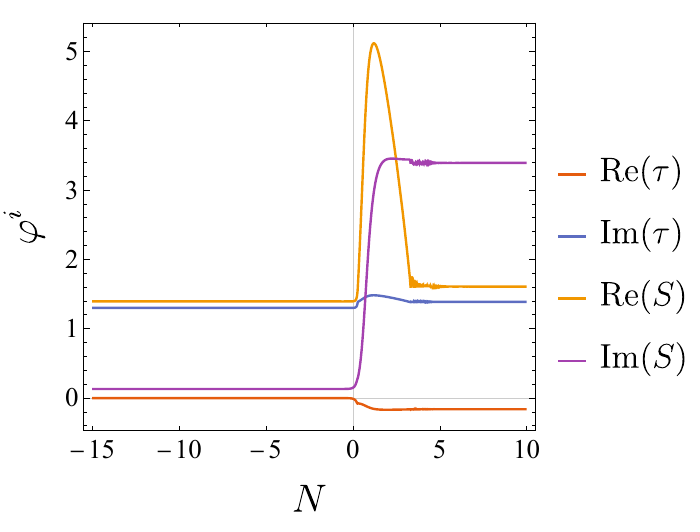}
    }
  \hspace{0.5cm}
   \subfloat{%
      \includegraphics[width=0.37\textwidth]{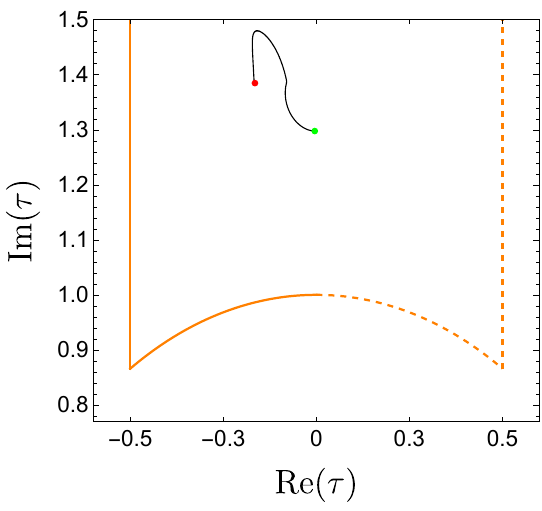}
    }
    \caption{\label{fig:FieldsCompleteEv}
    Evolution of the fields $\varphi_i$. 
    The left panel shows the values of the fields as a function of the number of e-folds 
$N$. The right panel presents the parametric plot of the K\"ahler modulus $\tau$. 
The green dot marks the saddle point, while the red dot represents the \ac{adS} minimum. It can be observed that the fields oscillate around the minimum until they stabilise.}
\end{figure}

Remarkably, as already noted, the fields evolve towards one of the \ac{adS} minima of the full potential. 
This implies that the current accelerated expansion is only a \emph{transient} phenomenon. 
To numerically study the complete trajectory, we uplift the potential by a constant term such that the minimum is shifted to zero, ensuring positivity of the potential throughout the evolution:
\begin{equation}\label{eq:tool}
    V_{\rm up} ~=~ V + \varepsilon \, ,
\end{equation}
with $\varepsilon=9.397\x10^{-122} %M_\mathrm{Pl}^4
$. 
Evolving the system with this uplift, we observe the fields rolling towards the nearby \ac{adS} minimum, oscillating around it as shown in the left panel of \Cref{fig:FieldsCompleteEv}\footnote{See e.g.~\cite{Sen:2021wld,Adil:2023ara, Menci:2024rbq, Wang:2024hwd} for cosmological constraints on models with a negative cosmological constant.}. 
From the numerical solution we extract the precise location of the minimum, its potential value, and the Hessian eigenvalues, summarised in \Cref{table:parametersAdS}. 
The right panel of \Cref{fig:FieldsCompleteEv} illustrates the full trajectory of the K\"ahler modulus, with the \ac{adS} vacuum located at $\tau = -0.1650 + 1.385\,\I$, lying within the fundamental domain. 

As discussed in \Cref{sec:susyadS}, all \ac{adS} minima we identify are supersymmetric up to numerical precision. At the endpoint minimum reached dynamically, we compute the $F$-terms, proportional to $D_\Psi W$, as well as the superpotential $W$ (before including the constant $A$). Our results read
\begin{equation}
\begin{split}
    \vev{D_\tau W} &~=~ 1.24\x 10^{-19}-5.25\x 10^{-20} \, \I \,, \\
    \vev{D_S W} &~=~ -6.10\x 10^{-20}-4.28\x 10^{-20} \, \I \,, \\
    \vev{W} &~=~ 2.04\x 10^{-5} -7.02\x 10^{-6} \, \I \, .
\end{split}
\end{equation}
Moreover, the ratio between the first and second terms in~\Cref{eq:Vsugra}, 
$K^{A\bar{B}} D_A W D_{\bar{B}} \overline{W}$ and $3|W|^2$, 
is $1.30 \x 10^{-28}$, again confirming supersymmetry to numerical accuracy. 

\begin{figure}[H]
    \centering
    \includegraphics[width=0.65\linewidth]{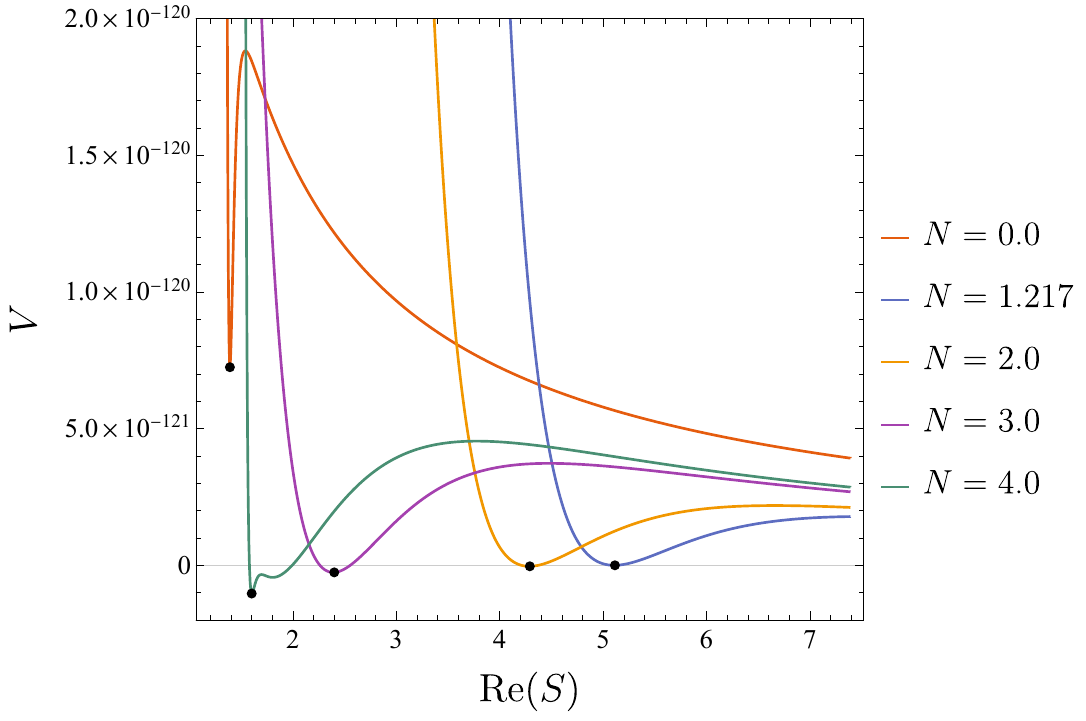}
    \caption{\label{fig:Vdilaton}
    Scalar potential $V$ as a function of $\re(S)$ for different e-fold values $N$. The black dots indicate the evolution of $\re(S)$ as $N$ increases, from \ac{dS} at $N=0$ (today) to \ac{adS} at $N=4$ (distant future). The field remains consistently at a local minimum along this direction throughout its evolution.}
\end{figure}

At first glance, the fact that the system evolves towards a vacuum rather than a runaway, as in~\cite{Olguin-Trejo:2018zun,Bhattacharya:2024kxp}, may seem surprising. 
However, this behaviour can be understood from the structure of the potential. 
Indeed, only $\re(S)$ admits a genuine runaway. 
The axionic components, $\re(\tau)$ and $\im(S)$, are stabilized by the expected periodic potentials, while $V$ diverges as $\im(\tau)\to\infty$. 
Although the $\re(S)$ direction exhibits a runaway behaviour, the dilaton initially sits at a local minimum and, throughout the entire cosmological evolution, it closely follows this minimum. 
We illustrate the dynamics of $\re(S)$ explicitly in \Cref{fig:Vdilaton}.

In summary, in the class of models we discuss, the present accelerated expansion is not eternal: the universe is ultimately destined to settle into a supersymmetric \ac{adS} vacuum. 
  
\begin{table}[t]
\centering
\begin{tabular}{|c|c|c|}
\hline
\rowcolor{gray!20}
\multicolumn{3}{|c|}{\textbf{Moduli \acp{VEV}}} \\
\hline
$\Psi^I$ & $\re\,\langle \Psi^I \rangle$ & $\im\,\langle \Psi^I \rangle$ \\
\hline
$\tau$ & $-0.1650$ & $1.385$ \\
\hline
$S$ & $1.605$ & $3.389$ \\
\hline
\rowcolor{gray!20}
\multicolumn{3}{|c|}{\textbf{Mass eigenstates}} \\
\hline
$\psi^I \sim \Psi^I$ & $m^2_{\re(\psi^I)}$ & $m^2_{\im(\psi^I)}$ \\ 
\hline
$\psi^\tau$  & $1.925\x 10^{-116}$ & $1.920\x 10^{-116}$ \\ 
\hline
$\psi^S$ & $5.292\x 10^{-118}$ & $5.206\x 10^{-118}$ \\ 
\hline
\rowcolor{gray!15}
\multicolumn{3}{|c|}{Potential value: $V=-1.052\x 10^{-121}$} \\
\hline
\end{tabular}
\caption{\label{table:parametersAdS}
AdS minimum of the potential. With the uplift in \Cref{eq:tool}, the fields evolve until reaching this stable vacuum. }
\end{table}

\enlargethispage{\baselineskip}

%% file: appendixA.tex
\section{A concrete orbifold model}
\label{App:concrete_model}

Our model arises from the heterotic string compactified on the orbifold $\Z{6}-\mathrm{II}$.
The model is fully specified by its twist $v$, shift $V$ and Wilson loops $A_{5}$ and $A_{6}$ given by
\footnotesize
\begin{subequations}\label{eq:shiftv}
\begin{eqnarray}
    v&=& \frac{1}{6}(0, 1, 2, -3),\\
    V &=&\left(-\frac{5}{12}, -\frac{5}{12},-\frac{5}{12}, -\frac{5}{12}, \frac{1}{12}, \frac{1}{12}, \frac{1}{12}, 
    \frac{1}{12}\right),\; \left(-\frac{5}{12}, -\frac{1}{4}, -\frac{1}{4}, -\frac{1}{12},\frac{1}{12}, \frac{1}{12}, 
    \frac{1}{12}, \frac{1}{12}\right) \, ,\\
    A_5  &=& \left( -\frac{5}{4},   \frac{3}{4},   \frac{5}{4},   \frac{7}{4},  -\frac{3}{4},   \frac{1}{4},   \frac{3}{4},   
    \frac{5}{4}\right),  \left( -\frac{1}{2},     1,     1,  -\frac{3}{2},    -1,     0,     1,     1\right) \, ,  \\
    A_6  &=&  \left( -\frac{3}{2},  -\frac{1}{2},     1,     1,   \frac{1}{2},     0,  -\frac{1}{2},     1 \right),  
    \left( -\frac{1}{4},   \frac{3}{4},   \frac{3}{4},   \frac{5}{4},  -\frac{1}{4},   \frac{7}{4},   \frac{1}{4},   
    \frac{7}{4}\right)\, . 
\end{eqnarray}    
\end{subequations}
\normalsize
By using the \texttt{Orbifolder}~\cite{Nilles:2011aj}, we obtain the unbroken \ac{4D} gauge group  
\begin{equation}
    G ~=~ \SU{3}_C\times\SU{2}_L\times\U{1}_Y \times \left(\SU{3}^3 \times \U{1}^6 \right)_{\text{hidden}}\,.
\end{equation}
We note that the non-Abelian factors $\SU3^3_\text{hidden}$ of the gauge hidden sector easily give rise to the gaugino condensates in \Cref{eq:Wgc}. We have explicitly verified the decoupling of $\SU3_\text{hidden}$-charged matter states and computed the beta-function coefficients before and after decoupling, $b_1$ and $b_1^0$, as well as $\delta_\text{GS}$ for one of these gauge factors. With this stringy motivation, we have assumed other suitable values for the corresponding coefficients of the second gaugino condensate to favour the structure of the potential.

The $\Z6-\text{II}$ $(1,1)$ orbifold can be factorised as $\mathds T^2/\Z6\otimes\mathds T^2/\Z3\otimes\mathds T^2/\Z2$. On the other hand, from \Cref{eq:shiftv}, we see that only the third torus is equipped with non-trivial Wilson lines. Hence, as discussed in~\Cref{sec:Modular}, the finite modular symmetry related to the $\mathds T^2/\Z2$ orbifold sector is broken, and, for simplicity, we can focus solely on the symmetries and dynamics of the moduli related to the $\mathds T^2/\Z3$ orbifold sector.

An appealing feature of $\mathds T^2/\Z{3}$ is its eclectic flavour symmetry group~\cite{Nilles:2020nnc}, consisting of a traditional flavour symmetry $\Delta (54)$, a finite modular symmetry $T'$ and a $\Z{9}^R$ symmetry, emerging from the 
outer automorphisms of its Narain lattice~\cite{Baur:2024qzo}. 
A proper discussion of the flavour problem requires considering both 
traditional and modular flavour symmetries, as they significantly 
constrain the K\"ahler potential and superpotential 
(see~\cite{Baur:2022hma} for an explicit study of flavour phenomenology with this 
eclectic structure). The associated matter spectrum, including charges under this group, is summarized in \Cref{tab:flavourcharges}.
The full expression of the K\"ahler potential, including
moduli and matter terms, is then 
\begin{equation}
    \label{eq:KahlerCompleto}
     K=-\log\left[S+\overline{S} - \frac{1}{8\pi^{2}}\delta_{\text{GS}}\log\left(\I\overline{\tau}- \I \tau\right) \right] 
     -\log\left(\I\overline{\tau}- \I \tau\right) + \left|\Phi_{\alpha}\right|^2 \left(\I\overline{\tau}-
     \I \tau\right)^{n_{\alpha}},
 \end{equation} 
which transforms covariantly under the modular symmetry, as in \Cref{eq:KandWtransformation}. 
In the \ac{LEEFT}, the last term is suppressed due to the relatively small \acp{VEV} of $\Phi_\alpha$ associated with the cancellation of the \ac{FI}-term (see \Cref{sec:SUSYmodStab}). Further, the assumption of a large volume regime and the restriction to negative fractional modular weights contribute to this suppression. The latter restriction is related to the fact that {\it massless} states with weights $n_\alpha=\nicefrac23,\nicefrac53$ are rather rare in heterotic orbifold compactifications as they exhibit oscillator excitations.
\begin{table}[]
    \centering
    \begin{tabular}{|c|c|c|c|c|}
    \hline
        sector & string state & $T'$ irrep & $\Delta(54)$ irrep & $\Z{9}^{R}$ charge\\
        \hline 
       \multirow{2}{*}{untwisted} & $\Phi^{(0)}$ & \multirow{2}{*}{$\rep{1}$ }& $\rep{1}$ & 0 \\
       \cline{2-2}\cline{4-5}
       & $\Phi^{(-1)}$ &  & $\rep{1}'$ & 3 \\ 
       \hline
       \multirow{2}{*}{first twisted} & $\Phi^{(-\nicefrac23)}$ & \multirow{2}{*}{$\rep{2}'\oplus\rep{1}$} & $\rep{3}_{2}$ & 1 \\
       \cline{2-2}\cline{4-5}
       & $\Phi^{(-\nicefrac53)}$ &  & $\rep{3}_{1}$ & $-2$ \\ 
       \hline
       \multirow{4}{*}{second twisted} & $\Phi^{(-\nicefrac13)}$ & \multirow{4}{*}{$\rep{2}''\oplus\rep{1}$}  & $\crep{3}_{1}$ & 2 \\
       \cline{2-2}\cline{4-5}
       & $\Phi^{(\nicefrac23)}$ &  & $\crep{3}_{2}$ & 5 \\
       \cline{2-2}\cline{4-5}
       & $\Phi^{(-\nicefrac43)}$ &  & $\crep{3}_{2}$ & $-1$ \\
       \cline{2-2}\cline{4-5} 
       & $\Phi^{(\nicefrac53)}$ &  & $\crep{3}_{1}$ & $-1$ \\
       \hline
    \end{tabular}
    \caption{\label{tab:flavourcharges} Spectrum of the $\mathds{T}^2/\Z3$ orbifold by sector and its irreps under the full flavour eclectic symmetry 
    $T'\cup\Delta(54)\cup\Z{9}^{R} $~\cite{Baur:2024qzo}.}
\end{table} 

Now we turn our attention to the Yukawa sector of the effective supergravity theory. For the computation of the superpotential we rely on the selection rules~\cite{Kobayashi:2011cw} among strings in orbifolds, as implemented in the \texttt{Orbifolder}. 
Considering the relevant terms up to order five in the model presented above, we obtain (see also~\cite{Nilles:2020kgo})
\begin{multline}
\label{eq:Wyuk1}
    W_{\mathrm{Yuk}} ~\supset~ \lambda\,\chi_{1}\,\chi_{2}\left[-\frac{\hat{Y}_{1}(\tau)}{\sqrt{2}}\left( \Phi_{1,3}\Phi_{2,2}\Phi_{3,1} + \Phi_{1,3}\Phi_{2,1}\Phi_{3,2} + \Phi_{1,2}\Phi_{2,3}\Phi_{3,1} \right.\right.\\ 
    + \left.\left.\Phi_{1,1}\Phi_{2,3}\Phi_{3,2}+ \Phi_{1,2}\Phi_{2,1}\Phi_{3,3} + \Phi_{1,1}\Phi_{2,2}\Phi_{3,3}  \right)\right.   \\ 
    \left.\right. 
    \left. +\hat{Y}_{2}(\tau)\left( \Phi_{1,1}\Phi_{2,1}\Phi_{3,1} +\Phi_{1,2}\Phi_{2,2}\Phi_{3,2} +\Phi_{1,3}\Phi_{2,3}\Phi_{3,3} \right)\right]\,, 
\end{multline}
where $\lambda$ is a constant to be fixed, $\chi_{\beta}$ are (gauge and flavour) singlets, $\Phi_{\alpha,\,p}$ are the $p$-th component of the $\alpha$-th triplet with modular weight $-\nicefrac23$ and $\hat{Y}_{i}$ are the modular forms defined in \Cref{eq:modularforms}. 
As mentioned before, the fields $\Phi_{\alpha,\,p}$ and $\chi_\beta$  acquire \acp{VEV} to cancel the \ac{FI}-term. Thus, defining 
\begin{subequations}
\begin{align}
\begin{split}
     \tilde{\lambda}_{1} &~:={}~ \lambda\vev{\chi_{1}}\vev{\chi_{2}}\left[ \vev{\Phi_{1,3}}\vev{\Phi_{2,2}}\vev{\Phi_{3,1}} + \vev{\Phi_{1,3}}\vev{\Phi_{2,1}}\vev{\Phi_{3,2}} + \vev{\Phi_{1,2}}\vev{\Phi_{2,3}}\vev{\Phi_{3,1}} \right.\\
     &\quad\,\,\, \left. + \vev{\Phi_{1,1}}\vev{\Phi_{2,3}}\vev{\Phi_{3,2}} + \vev{\Phi_{1,2}}\vev{\Phi_{2,1}}\vev{\Phi_{3,3}} + \vev{\Phi_{1,1}}\vev{\Phi_{2,2}}\vev{\Phi_{3,3}} \right]\,,
\end{split}\\
     \tilde{\lambda}_{2} &~:={}~ \lambda\vev{\chi_{1}}\vev{\chi_{2}}\left[ \vev{\Phi_{1,1}}\vev{\Phi_{2,1}}\vev{\Phi_{3,1}} + \vev{\Phi_{1,2}}\vev{\Phi_{2,2}}\vev{\Phi_{3,2}} + \vev{\Phi_{1,3}}\vev{\Phi_{2,3}}\vev{\Phi_{3,3}} \right]\,,
\end{align}
\end{subequations}
\Cref{eq:Wyuk1} takes the simplified form~\eqref{eq:Wyuk}, i.e.
\[
W_{\text{Yuk}} ~=~ -\frac{1}{\sqrt{2}}\tilde{\lambda}_1 \hat{Y}_{1}(\tau) + \tilde{\lambda}_2 \, \hat{Y}_{2}(\tau)  + \cdots
\, .\]

At leading order, this approximation includes $11$
out of a set of $\mathcal{O}(100)$ matter fields.
All additional states appear at higher orders and 
are hence further suppressed. This implies that their
contributions to $W_\text{Yuk}$ can be ignored at
first order.